\documentclass[%
 reprint,
nofootinbib,
 amsmath,amssymb,
 aps,
prd,
]{revtex4-2}
\usepackage{siunitx}
\usepackage[hidelinks,colorlinks=true,linkcolor=blue,citecolor=blue]{hyperref}
\usepackage{natbib}
\usepackage{CJK}
\usepackage{graphicx}
\usepackage{dcolumn}
\usepackage{bm}
\usepackage[dvipsnames]{xcolor}

\def\eg{\textit{e.g.}}
\def\ie{\textit{i.e.}}
\def\holodeck{\texttt{holodeck}}

\begin{document}


\title{Bridging the micro-Hz gravitational wave gap via Doppler tracking with the\\  Uranus Orbiter and Probe Mission:\\ Massive black hole binaries, early universe signals and ultra-light dark matter}


\author{Lorenz Zwick$^{1}$}
\email{lorenz.zwick@nbi.ku.dk}
\author{Deniz Soyuer$^{2}$}
\author{Daniel J. D'Orazio$^{1,3}$}
\author{David O'Neill$^{1}$}
\author{Andrea Derdzinski$^{4,5}$}
\author{Prasenjit Saha$^{6}$}
\author{Diego Blas$^{7,8}$}
\author{Alexander C. Jenkins$^{9}$}
\altaffiliation{Present address: Kavli Institute for Cosmology, University of Cambridge, Madingley Road, Cambridge CB3 0HA, UK and DAMTP, University of Cambridge, Wilberforce Road, Cambridge CB3 0WA, UK}
\author{Luke Zoltan Kelley$^{10}$}
\affiliation{%
\vspace{.2cm} $^{1}$Niels Bohr International Academy, The Niels Bohr Institute, Blegdamsvej 17, DK-2100, Copenhagen, Denmark\\
 $^{2}$Department of Astrophysics, University of Zurich, Winterthurerstrasse 190, 8057 Zurich, Switzerland\\
 $^{3}$Space Telescope Science Institute, 3700 San Martin Drive, Baltimore, MD 21218, USA \\
 $^{4}$Department of Life and Physical Sciences, Fisk University, 1000 17th Avenue N., Nashville, TN 37208, USA\\
 $^{5}$Department of Physics \& Astronomy, Vanderbilt University, 2301 Vanderbilt Place, Nashville, TN 37235, USA\\
 $^{6}$Physik-Institut, Universität Zürich, Winterthurerstrasse 190, CH-8057 Zürich, Switzerland\\
 $^{7}$Institut de Fisica d’Altes Energies (IFAE), The Barcelona Institute of Science and Technology, Campus UAB, 08193 Bellaterra (Barcelona), Spain\\
 $^{8}$Instituci\'o Catalana de Recerca i Estudis Avan\c cats (ICREA), Passeig Llu\'is Companys 23, 08010 Barcelona, Spain\\
 $^{9}$Department of Physics and Astronomy, University College London, London WC1E 6BT, UK\\
 $^{10}$Department of Astronomy, University of California, Berkeley, California 94720, USA} 


%


\date{\today}

\begin{abstract}

 With the recent announcement by NASA's \textit{Planetary Science and Astrobiology Decadal Survey 2023-2032}, a priority flagship mission to the planet Uranus is anticipated. Here, we explore the prospects of using the mission's radio Doppler tracking equipment to detect gravitational waves (GWs) and other analogous signals related to dark matter (DM) over the duration of its interplanetary cruise. We develop a methodology to stack tracking data and account for time varying detector geometry, thereby constructing the sensitivity of the mission to GWs over the wide frequency range of $3\times 10^{-9}$ Hz to $10^{-4}$ Hz. We find that the mission has the potential to fill the gap between pulsar timing and space-based-interferometry GW observatories. If improvements in reducing \textit{Cassini} era noise by a factor of $\mathcal{O}(10)$ are implemented, we forecast the detection of $\mathcal{\mathcal{O}}(\rm{few})$ individual massive black hole binaries using two independent population models. Additionally, we determine the mission's sensitivity to both astrophysical and primordial stochastic gravitational wave backgrounds, as well as its capacity to test, or even confirm via detection, ultralight DM models. In all these cases, the tracking of the spacecraft over its interplanetary cruise would enable coverage of unexplored regions of parameter space, where signals from new phenomena in our Universe may be lurking.
\end{abstract}

\maketitle


%

\section{Introduction} 
\label{S:Introduction}
\begin{figure}
    \centering
    \includegraphics[width = \columnwidth]{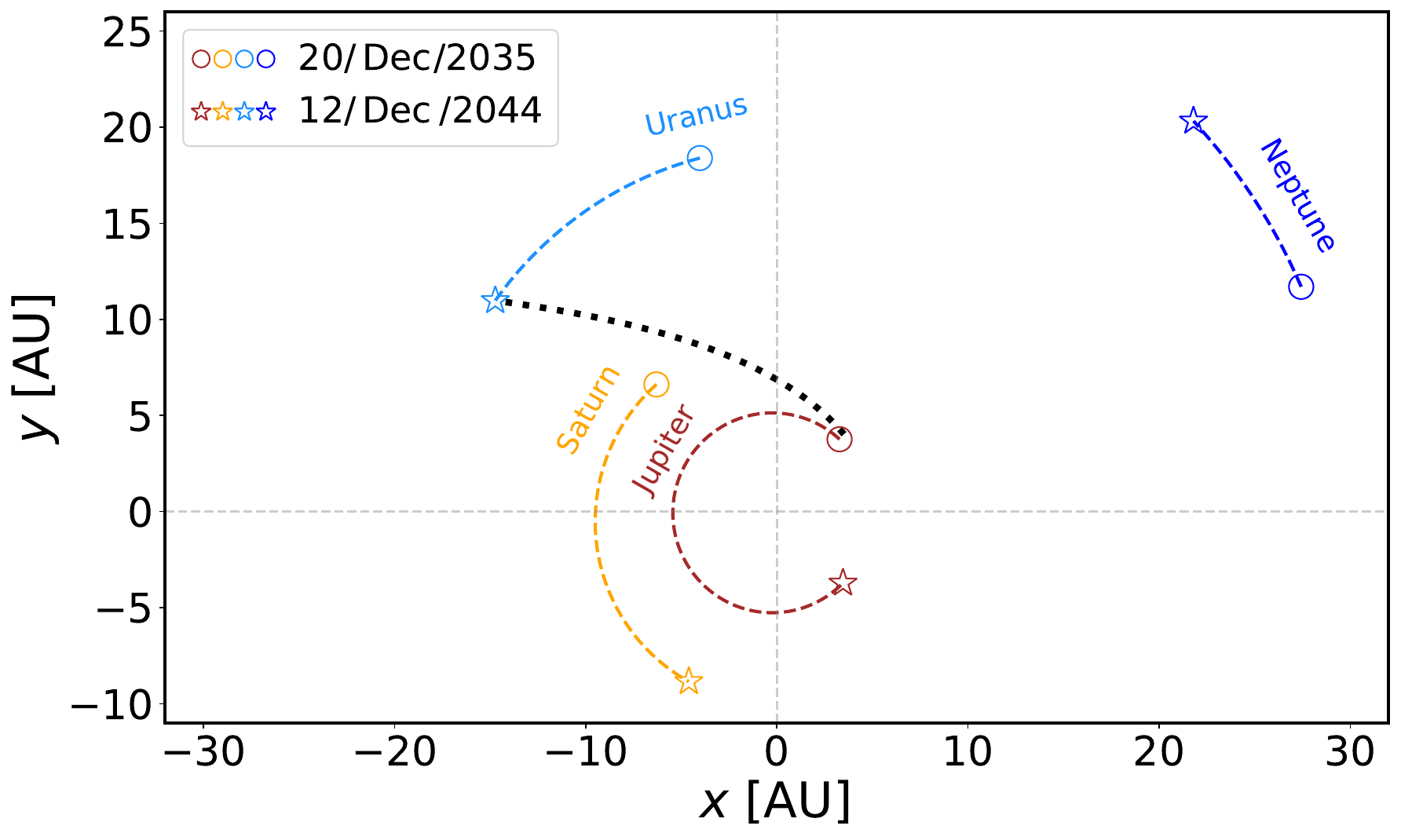}
    \caption{One of the many proposed mission plans of a prospective Uranus mission that involves a Jupiter Gravity Assist and a subsequent cruise to Uranus. 
     This specific mission timeline corresponds to the shortest proposed cruise time and is available in the public \href{https://smd-cms.nasa.gov/wp-content/uploads/2023/10/uranus-orbiter-and-probe.pdf}{mission document}.}
    \label{fig:traj}
\end{figure}

Almost half a century after the launch of the \textit{Voyager 2} space probe, the prospect of a new visit to the ice giant Uranus is finally crystallizing again: A flagship mission to the planet Uranus (hereafter referred to as ``UOP'', standing for Uranus Orbiter and Probe) has been declared a priority, according to the announcement by \textit{NASA’s Planetary Science and
Astrobiology Decadal Survey 2023-2032}.\footnote{Survey provided in: \href{https://doi.org/10.17226/26522}{Origins, Worlds, and Life: A Decadal Strategy for Planetary Science and Astrobiology 2023-2032 (2022)}} The announcement is timely; numerous publications have underlined the rich potential of such a mission in terms of planetary science over the past few years~\citep{hofstadter, fletcher2, fletcher,helled, quest, simon2020, kollmann, girija, euro}. 
Yet, any mission to the outer Solar system must undergo a long cruise in interplanetary space before reaching its destination. According to a notional, short duration trajectory of the UOP (see Fig.~\ref{fig:traj}), just the transfer between Jupiter and Uranus is expected to require around 9 years, during which the planetary science yield is virtually non-existent.

Several works have highlighted how the UOP's long cruise may instead provide unique opportunities for \textit{non-Uranian science}. Prospects include the possible measurement of the local dark matter content in the Solar system~\citep{2006sereno,zwick2022}, as well as the localisation of the hypothetical Planet 9 in the sky~\citep{bucko2023}. The opportunity to take such measurements is afforded by the presence of a radio link between the spacecraft and the Earth, which would allow one to closely monitor changes in the spacecraft's velocity and reconstruct deviations from the expected mission trajectory. Above all, the most timely application of interplanetary radio tracking is arguably the detection of gravitational wave (GW) signals. The idea of detecting GWs in the Doppler tracking data of planetary missions has a rich history and has been attempted previously with 
\textit{Pioneer 11}~\citep{pioneer11}, the \textit{Galileo}--\textit{Ulysses}--\textit{Mars} \textit{Observer} coincidence experiment~\citep{galileo, ulysses}, and \textit{Cassini}~\citep{cassini92, godtierpaper,2003ApJ...599..806A,2006armstrong}, albeit without any successful detection candidates due to insufficient sensitivity. As first highlighted in Ref.~\cite{soyuer2021}, the prospective UOP offers an exceptional opportunity compared to the previous attempts due to a longer-duration spacecraft cruise, a longer detector arm, and potentially improved radio technology.\footnote{Recently, there was another proposal to use tracking of satellites to Mars to study GWs of frequencies from $10^{-4}$ Hz to $10^{-1}$ Hz~\citep{2024arXiv240206096B}. As compared to our proposal, these constraints are too weak to be relevant for known sources, and they will be largely superseded in the future by LISA~\citep{redbook}.}

In this work, we consider how the collection of tracking data over the entirety of the UOP's interplanetary cruise can increase the sensitivity of the mission to GWs from many sources, allowing us to probe the yet poorly covered micro-Hz regime. Additionally, we outline the mission's capability to greatly expand the range of ultra-light dark matter (ULDM) models~\citep{Hui17,2017PhRvL.118z1102B} that can be directly detected, including the possibility of a first detection of DM in the Solar System. The key for these results is a method developed here to stack individual short-duration tracking runs {over a slowly changing detector geometry, in order to take advantage of the long detector arm constituted by the Earth-spacecraft system}. The mission's GW frequency sensitivity band and DM mass sensitivity are in principle only limited by the total observation time and the data cadence, in analogy to Pulsar Timing Array analyses for nano-Hz GW detection~\citep[hereafter PTA, see, \eg,][]{1990forster,2008sesana,2010hobbs,EPTA-GWB:2023,NANOG-GWB:2023}.

In summary, we show that, beyond its Uranian science goals, the UOP could double as a sensitive GW detector in the micro-Hz band, probing the astrophysics of massive black holes or signals from the early Universe, and as a dark matter detector for models not accessible today by any other probe. In principle, this should be achievable without any sacrifice to planetary science goals by tracking the mission consistently over its interplanetary cruise. We also remark that the mission is expected to fly in concomitance with the Laser Interferometer Space Antenna (LISA~\cite{redbook}). The opportunities for complementarity between these two missions were briefly discussed in Ref.~\cite{soyuer2021}. They deserve a more thorough analysis, which we leave for future work.

This paper is structured as follows: In Section \ref{S:doppler} we introduce the basics of GW and ULDM detection via Doppler tracking and detail our strategy to track signals over the entire mission duration. In Section \ref{sec:senscurves}, we construct GW sensitivity curves and define three possible mission configurations with increasing levels of technological development. In Section \ref{S:smbh} we estimate the number of detections of individual massive black hole binaries with two independent population models, as well as the associated stochastic gravitational wave background. 
Section \ref{sec:earlyuniverse} and Section \ref{sec:uldm} include our forecast constraints on GW signals from the early universe, as well as various DM models, respectively. Lastly, we discuss the prospects to further improve the sensitivity of the mission in Section \ref{sec:noise} and present our concluding remarks in Section \ref{sec:conclusion}.

\section{Doppler Tracking of GWs and other signals} 
\label{S:doppler}
\subsection{Response of Doppler tracking to a passing GW}
 The passage of a GW between the Earth and the spacecraft modifies the propagation time of photons, inducing small timing shifts in the tracking system. These shifts are best described by defining the dimensionless fractional frequency fluctuation of a two-way Doppler system:
\begin{equation}
    y_2(t) = {\Delta \nu}/{\nu_0}, \label{eq:Doppler1}
\end{equation}
where $\nu_0$ is the tracking system's carrier frequency, most often in the Ka-band ($\sim$30\,GHz) or the X-band ($\sim$8\,GHz), while $\Delta \nu$ is the measured Doppler shift with respect to the carrier frequency.
Following Ref.~\cite{2006armstrong}, we can write down an explicit formula for the frequency fluctuation due to a passing GW:
\begin{align}
\nonumber
    y_2^{\scriptscriptstyle\rm{GW}}(t) = \frac{\mu(t) -1}{2} \bar{\Psi}(t) ~-~& \mu(t) \bar{\Psi}\left(t - \frac{\mu(t) + 1}{2} T_2(t) \right)\\ 
    ~+~& \frac{\mu(t)+1}{2} \bar{\Psi}(t - T_2(t)),
    \label{eq:y2}
\end{align}
where $T_2$ is the two-way light travel time between the Earth and the spacecraft, $\mu = \hat{\boldsymbol{k}} \cdot \hat{\boldsymbol{n}}$ is the projection of the unit wave vector $\hat{\boldsymbol{k}}$ of the GW onto the unit vector connecting the Earth and the spacecraft $\hat{\boldsymbol{n}}$, and $\bar{\Psi}$ is the projection of the GW amplitude
onto the Doppler link~\citep{psi}. The latter is given by $\bar{\Psi}(t) = (\hat{\boldsymbol{n}} \cdot \mathrm{\bf{h}}(t) \cdot \hat{\boldsymbol{n}})/(1 - \mu^2)$,
where $\mathrm{\bf{h}}(t) = \mathrm{{h}}_+(t) \, \mathrm{\bf{e}}_+ + \,\mathrm{{h}}_\times(t) \, \mathrm{\bf{e}}_\times$ is the GW amplitude (\ie, the strain) and $\mathrm{\bf{e}}_{+,\times}$ are the usual ``plus'' and ``cross'' polarization states of a transverse, traceless plane GW\,\citep[][]{1956pirani,1959bondi}.

{While Eq.~\eqref{eq:y2} can be used to describe any incident GW, we are primarily interested in two particular types of signal.
The first is coherent, monochromatic GWs of the form
\begin{subequations}
\begin{align}
\label{eq:h}
    &\mathrm{h}_+(t)= A \frac{1 + \cos^2 \iota}{2}\cos(f_{\scriptscriptstyle\rm GW} t + \phi_0), 
    \\ &\mathrm{h}_\times(t) = A \cos \iota \sin(f_{\scriptscriptstyle\rm GW} t + \phi_0)\,,
\end{align}
\end{subequations}
that are emitted by inspiralling massive black hole binaries (where $f_{\scriptscriptstyle\rm GW}$ is the frequency of the GW, $A$ its amplitude, $\iota$ is the binary's inclination, and $\phi_0$ is an initial phase~\citep[see, \eg,][]{gravitation,moore,maggiore}); these are the most promising astrophysical source of GWs in the nano-Hz to micro-Hz frequency band, and will be discussed in detail in Section \ref{S:smbh}.}

{The second family of signals we consider are stochastic GW backgrounds (SGWBs) sourced by the incoherent superposition of many astrophysical sources (see, \eg\,\citep{Sesana:2019vho} for examples in the micro-Hz band), or via fundamentally stochastic processes in the early Universe~\citep{Caprini18}.
SGWBs are modeled by considering the sky- and polarization-averaged power spectrum of the observable given in Eq.~\eqref{eq:y2}, $S_{y_2}(f)$.
Following~\citep{Estabrook75}, this quantity can be expressed as $S_{y_2}(f)=\bar R_2(f)S_{\scriptscriptstyle\rm GW}(f)$, where $S_{\scriptscriptstyle\rm GW}(f)$ is the GW power spectrum, and the transfer function $\bar R_2(f)$ can be found in Ref.~\citep{2003ApJ...599..806A}. In the limit of low frequencies ($2\pi f T_2\ll 1$), it reads simply $\bar R_2(f)\approx (8 / 15)\left(\pi f T_2\right)^2$, while at higher frequencies it displays some oscillatory behavior.
Rather than the GW power spectrum itself, it is often convenient to express constraints on the SGWB in terms of its energy density spectrum, normalized to the critical cosmological energy density,
\begin{equation}
\Omega_{\scriptscriptstyle\rm GW}(f) \equiv \frac{8\pi G}{3 H_0^2} \frac{\mathrm d \rho_{\scriptscriptstyle\rm GW}}{\mathrm d \log f } =\frac{8 \pi^2 f^3}{3H_0^2}S_{\scriptscriptstyle\rm GW}(f).
\end{equation}
We will consider the Hubble parameter to be $H_0\approx 67\,$km/s/Mpc~\citep{Planck:2018vyg}.}

\subsection{Response of Doppler tracking to ultra-light dark matter}\label{sec:ULDMsignal}
 If dark matter (DM) is composed of particles of masses $m_{\scriptscriptstyle\rm DM}\lesssim 1$~eV/$c^2$ they will be distributed in the Milky Way at inter-distances smaller than their de Broglie wavelength (associated with the `size' of the particle). As a result, DM can be approximated in this range of masses as a classical non-relativistic massive field $\phi_{\scriptscriptstyle\rm DM}$~\citep{Hui17}. From the virialised properties of DM in the Milky Way, this field will be distributed in coherent patches of characteristic length \citep{ferreira21}:
\begin{equation}
    l\approx 0.9 \times 10^{6} \mathrm{~AU}\left(\frac{10^{-3}}{\sigma_0}\right)\left(\frac{10^{-20} \mathrm{eV}/c^2}{m_{\scriptscriptstyle\rm DM}}\right),
\end{equation}
where $\sigma_0$ is the DM velocity dispersion, generated by virialization in the Milky Way. Its value extracted from data for the dark matter halo is $\sigma_0\approx 10^{-3}$~\cite{Lisanti:2016jxe}. In these regions, the DM field can be approximated as a monochromatic function oscillating at a frequency
\begin{equation}
    f_{\scriptscriptstyle\rm DM} \approx 2.4 \times 10^{-6}\left(\frac{m_{\scriptscriptstyle\rm DM}}{10^{-20} \mathrm{eV}/c^2}\right) \mathrm{Hz},
    \label{eq:m_to_Hz}
    \end{equation}
 coherent over $c^2/\sigma_0^2$ oscillations and with amplitude\footnote{Technically, the interference of the different classical waves that comprise the ULDM implies that the local density has a degree of randomness over this value, which corresponds to the average DM energy density~\citep{Foster:2017hbq}. We will assume that the DM energy density in the Solar System is 1/2 of the averaged one (expected to be of order $\rho_{\rm DM}\approx 0.3\,\rm{GeV/cm^3}$) to provide our estimates.} 
 \begin{equation}
      \phi_0\approx \sqrt{2\rho_{\rm DM}}/m_{\rm DM}.
      \label{eq:phiDM_rho}
 \end{equation}
 As a result, the gravitational potentials generated by the galactic DM (derived from solving the Poisson equation, which is quadratic in $\phi_{\scriptscriptstyle\rm DM}$) inherit the coherent features of the DM field, and oscillate with frequency $f_{\rm grav}\approx 2 f_{\scriptscriptstyle\rm DM}$ in these patches (cf. Eq.~\eqref{eq:psi_DM} below). When approaching the lowest viable masses, these DM models are known as fuzzy dark matter~\citep{Fuzzy} or ultra-light dark matter (ULDM)~\citep{Hui17}.

The Doppler tracking suggested above for GWs can be easily adapted to search for ULDM in different ways. 
First, notice that, at leading order, the fluctuations of the gravitational potential $g_{ij}\equiv -(1-2\psi)\delta_{ij}$ modify the relative frequency of the radio signal as~\citep{Khmelnitsky:2013lxt,2017PhRvL.118z1102B,Porayko:2018sfa,Unal:2022ooa}
\begin{equation}
    \Delta \nu/\nu_0= \psi(t)-\psi\left(t-T_2\right).\label{eq:fshiftDM}
\end{equation}
For ULDM, the fluctuating piece of the gravitational potential $\psi$ behaves (at leading order in $\sigma_0\approx 10^{-3}$, with corrections scaling as $\sigma_0^2$) as\footnote{As shown in Ref.~\cite{Kim:2023pkx}, the gravitational potential has also support at much lower frequencies, where it can be described as a stochastic background. We will study the influence of these fluctuations elsewhere.}
\begin{equation}
\begin{split}
\psi(t)\approx\psi_g(t)&\equiv \, 6 \cdot 10^{-16}\left(\frac{\rho_{\scriptscriptstyle\rm DM}}{0.3\, \mathrm{GeV}/ \mathrm{cm}^3}\right)\\&\left(\frac{10^{-23}\, \mathrm{eV}/c^2}{m_{\scriptscriptstyle\rm DM}}\right)^2\cos( 4\pi f_{\scriptscriptstyle\rm DM} t+ 2\alpha_0),
    \end{split}
\label{eq:psi_DM}
\end{equation}
where $\rho_{\scriptscriptstyle\rm DM}$ is the dark matter energy density and $\alpha_0$ is a random phase.
This immediately allows one to use the bounds on $y_2^{\scriptscriptstyle\mathrm{GW}}(t)$ to constrain the mass of dark matter candidates (it is enough to use \eqref{eq:fshiftDM} in \eqref{eq:Doppler1}, and connect it to \eqref{eq:y2}). To find the precise bound, it is important to realise that from the lack of polarization and directional dependence, a factor of $\sqrt{15}$ appears when comparing the bounds on rms of $h(t)$ vs $\psi(t)$~\citep{2017PhRvL.118z1102B} (see also Ref.~\cite{Khmelnitsky:2013lxt} for the comparison of these two quantities in the case of pulsar timing). 

Second, it is normally expected that DM is not completely `dark', and couples, though very weakly, to matter or light. As a result, the fluctuations of $\phi_{\rm DM}$ can affect the tracking data in new ways. For instance, new forces may affect the trajectory of gravitating probes or the properties of the tracking beam, see, \eg,~\citep{Miller:2022wxu,LIGOScientific:2021ffg,Fedderke:2022ptm,
Fukusumi:2023kqd,Oshima:2023csb,Nagano:2021kwx}. 
For our current study, we will focus on two simple, though representative possibilities of \emph{direct coupling} of DM to ordinary matter and light. In the first model, dark matter is coupled to the constituents of the detectors, the spacecraft, and the tracking beam through a \emph{universal coupling}, common to all particles of the standard model of particle physics. This universality is similar to the coupling of the metric $g_{\mu\nu}$ in general relativity. As a consequence, in this model the coupling of matter and light to gravity ($g_{\mu\nu}$) and dark matter ($\phi_{\rm DM}$) happens through the interaction with an effective metric $\bar g_{\mu\nu}=g_{\mu\nu}(1+2\alpha(\phi_{\scriptscriptstyle\rm DM}))$. In this situation, matter and light will move in geodesics of $\bar g_{\mu\nu}$, implying that the expression~\eqref{eq:fshiftDM} will still be valid, identifying\footnote{More specifically, at leading order in perturbations over the flat background, $\bar g_{ij}= -(1-2\psi-2\alpha(\phi_{\scriptscriptstyle\rm DM}))\delta_{ij}$.} $\psi\equiv \psi_g(t)+\alpha(\phi_{\scriptscriptstyle\rm DM})$. The function $\alpha(\phi)$ is model-dependent. To illustrate the potential of UOP to test DM models, we will choose $\alpha(\phi_{\scriptscriptstyle\rm DM})=\phi^2_{\scriptscriptstyle\rm DM}/\Lambda^2_2$, where $\Lambda_2$ parameterises the intensity of the coupling. The effect of ULDM is described in this case through~\eqref{eq:fshiftDM}, after the identification
\begin{equation}
\begin{split}
    \psi(t)\approx \left(1+40\left(\frac{10^{18}\rm{GeV}}{\Lambda_2}\right)^2\right)\,\psi_g(t),\label{eq:direct}
\end{split}
\end{equation}
where we used \eqref{eq:phiDM_rho} and \eqref{eq:psi_DM}.
To illustrate other possibilities opened from the coupling of the tracking (light) beam to DM, we will consider a model where DM is an axion-like particle~\citep{Chadha-Day:2021szb}. In this case, a coupling of the form 
\begin{equation}
    g_{a\gamma \gamma} \phi_{\rm DM}\vec E \cdot \vec B, \label{eq:couplingaxion}
\end{equation}
is expected, where $\vec E$ and $\vec B$ are the electric and magnetic fields respectively, and $g_{a\gamma \gamma}$ represents the intensity of the coupling. Following Ref.~\citep{Blas:2019qqp}, for the ULDM case one expects frequency oscillations in the tracking beam of order
\begin{equation}
    \frac{\Delta \nu}{\nu_0} \sim 10^{-16}\left(\frac{g_{a \gamma \gamma}}{10^{-10} \mathrm{GeV}^{-1}}\right)\left(\frac{ \mathrm{GHz}}{\nu/2\pi}\right) \sqrt{\frac{\rho_{\scriptscriptstyle\rm{DM}}}{0.3 \mathrm{GeV} / \mathrm{cm}^3}} \label{eq:shift}
\end{equation}
at frequency $f_{\scriptscriptstyle\rm DM}$, where we have normalised $g_{a \gamma \gamma}$ to a value representative of current sensitivity~\citep{AxionLimits,Graham:2015ouw}. In other words, surpassing this sensitivity for the range of masses of interest would likely be a way to test axion-like models currently unconstrained by other methods. 

The previous three effects, summarised by Eqs.~\eqref{eq:psi_DM},~\eqref{eq:direct} and~\eqref{eq:shift} are far from being comprehensive (for instance, we are not discussing the possibility of other possibilities for $\alpha(\phi_{\rm DM})$, beyond the quadratic case of \eqref{eq:direct}, or models where DM is a vector field, as in the case of dark photons). Still, they represent three families of ULDM models: universal gravitational coupling; direct coupling to matter; and direct coupling to light, which illustrate the potential of UOP to explore ULDM models. 


\subsection{Noise in a Doppler tracking system}
The response of a Doppler tracking system to GWs and other oscillatory signals must be compared to its intrinsic noise. Noise sources range from mechanical (caused, \eg, by flexibility in the antennae), to astrophysical (\eg, scintillation due to propagation through the interplanetary plasma). These are thoroughly compiled and discussed in the extensive review by Ref.~\cite{2006armstrong}, and further discussed in Section \ref{sec:noise}. More recently, various noise components have been analysed in Ref.~\cite{2024mcquinn}. In general, the total noise profile can be described in either the time domain or the frequency domain. Starting with the former case, {a ranging system's raw data consists of measurements of the radio link's instantaneous frequency. To reduce noise of individual measurements, many samples may be averaged over a characteristic timescale $\tau$. The variance of the remaining noise fluctuations for each averaged data-point is characterized by a quantity called the Allan deviation, $\sigma_{\rm A}$}, which is a function of the chosen averaging time $\tau$. {In essence, the Allan deviation measures the residual variance of the noise on the Doppler link given an averaging time $\tau$. Typically, $\sigma_{\rm A}$ deteriorates both at small values of $\tau$, where high frequency noise sources cannot be averaged out, and at large values of $\tau$, where systematic drifts in the ranging system start to become relevant. The Allan deviation is also related to the description of the system in Fourier space, where} noise is best described by its one-sided power spectrum $S_n$ (noise PSD), which is a function of frequency $f$. These two crucial quantities are related by the following equation:
\begin{align}
    \label{eq:allandevspectraldens}
    \sigma_{\rm A}(\tau) = \sqrt{ 4\int_{0}^{\infty} S_n(f) \frac{\sin(f\tau)^4}{(\pi f \tau)^2}\, {\rm d}f} \approx \sqrt{\frac{S_n(1/\tau)}{\tau}},
\end{align}
where the latter approximation can be verified numerically, and is exact in the case of white noise~\citep{godtierpaper}.
Eq.~\eqref{eq:allandevspectraldens} essentially states that the typical noise fluctuations given an averaging time $\tau$ are determined by the noise PSD at the reciprocal frequency $f \sim 1/\tau$.

\begin{figure}
    \centering
    \includegraphics[width = \columnwidth]{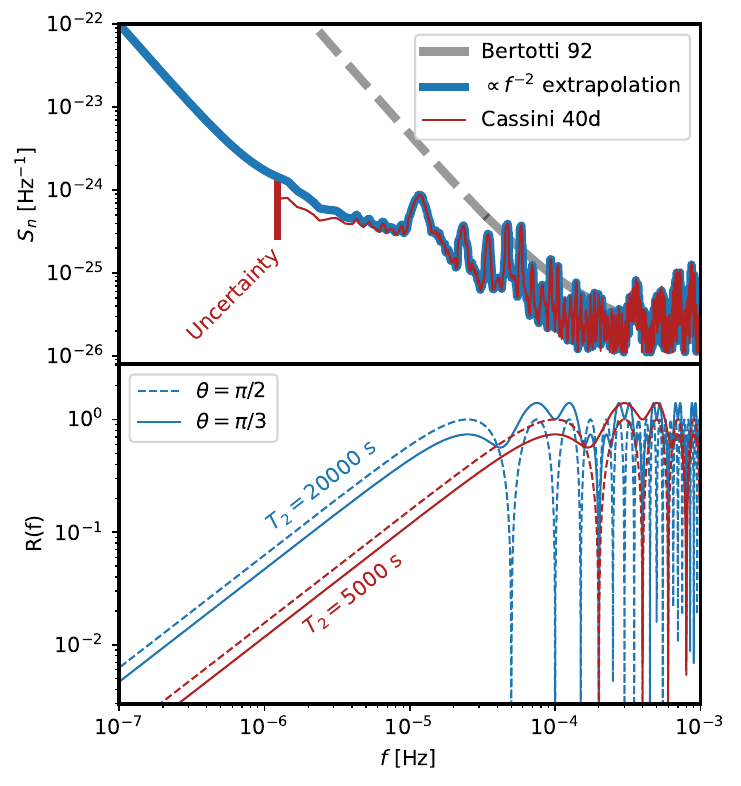}
    \caption{{Top panel: Observed Cassini PSD for a 40 day run~\citep{2006armstrong} in red, as well as the extrapolation to lower frequencies used in this work (blue, see text). The vertical red bar denotes the representative uncertainty for the low frequency part of the obsered PSD. The gray curve shows the noise PSD based on theoretical work by Ref.~\cite{ulysses} and analysis of the \textit{Ulysses} and \textit{Cassini} tracking data~\cite{cassini92}, which was shown to be too conservative. The extrapolation to frequencies lower than $10^{-4}$ Hz~\citep{godtierpaper} is shown as a dashed line. Bottom panel: Response transfer functions to GW for a fixed length Doppler tracking systems with a given incidence angle~\citep[][see also section \ref{sec:senscurves}]{2002tinto}. The strategy to integrate the response over a variable UOP trajectory is detailed in section \ref{subsec:stack}.}}
    \label{fig:cass_noise}
\end{figure}

{Simple parametrisations of noise PSDs for {\textit{Cassini}} era Doppler ranging systems are first discussed in Ref.~\cite{godtierpaper} and compared to extensive analysis of the \textit{Ulysses} tracking data~\citep{ulysses,cassini92}. The broad features of the expected PSD can be captured by a simple three power law structure. At frequencies higher than $f\sim 0.1$ Hz, the noise is dominated by thermal fluctuations and steeply increases as $\sim f^2$. This limit tends to constrain the minimum averaging time of tracking systems to $\tau \sim 10$ s, beyond which the Allan deviation rapidly deteriorates. A central region of optimal sensitivity spans between $\sim10^{-4}$ Hz and $\sim10^{-1}$ Hz and is characterised by a red spectrum $\sim f^{-1/2}$~\cite{cassini92}. The strong upturn in noise below $10^{-4}$~Hz (seen in Fig.~\ref{fig:cass_noise} in gray) was an attempt to model less constrained low-frequency noise sources, which range from uncertainties in the dynamical modelling of the spacecraft trajectory to the effects of atmospheric turbulence. They were conservatively estimated to generate a slope of $f^{-2}$, as had been originally observed in the \textit{Ulysses} tracking data. However, such estimates were purely phenomenological, and the consequences of flatter low-frequency slope for the noise, and the resulting improvements in sensitivity were already explored in~\cite{cassini92}. Approximately a decade later, the noise PSD of \textit{Cassini} down to frequencies of $\sim 10^{-5}$ Hz was estimated from first principles in Ref.~\cite{2002tinto}, in preparation for a 40 day tracking run performed in 2002. The latter prediction matched exceptionally well with the observed \textit{Cassini} noise, which can be found in Ref.~\cite{2003ApJ...599..806A,2006armstrong} and is plotted in Fig \ref{fig:cass_noise}.}

{For the purposes of this work, we will conservatively assume that the observed noise of the \textit{Cassini} Doppler tracking system is representative of modern interplanetary spacecraft within the relevant frequency range. To extend the noise PSD below the frequency of (40 days)$^{-1}$, we extrapolate the observed PSD with a phenomenological scaling of $f^{-2}$, as originally suggested by Ref.~\cite{cassini92}. The amplitude of the $f^{-2}$ component is chosen such that it matches the \textit{upper bound} of the representative uncertainty in the observed PSD~\citep[][see also Fig.~\ref{fig:cass_noise}]{2006armstrong}. We note here that none of the individual noise sources analysed in Refs.~\cite{2006armstrong} and~\cite{2024mcquinn} produces a noise component with a steeper slope. Therefore, our extrapolation can be considered \textit{conservative}, unless additional unexpected noise sources populate the $10^{-9}$ Hz to $10^{-6}$ Hz band. Here we also investigate how possible technological advancements in the last four decades may affect the sensitivity of the tracking system. To parametrise such improvements, we scale the entire noise PSD with a reference Allan deviation at 10 s, $\sigma_{\rm A}@10$ s:}
\begin{align}
    S_{\rm n}(f) \equiv \frac{\sigma_{\rm A}@10}{10^{-14}}S_{\rm Cass}^{\rm extr.}(f),
\end{align}
{where $S_{\rm Cass}^{\rm extr.}(f)$ is the observed \textit{Cassini} noise PSD with an additional $f^{-2}$ extrapolation. Note however, that the only relevant improvements for the purposes of our results are those specific to the micro-Hz range.} We discuss some of these advanced technological prospects in Section \ref{subsec:optical}. The effects of these choices will be further discussed in Section~\ref{sec:noise}.

\subsection{Strategy for the tracking of low-frequency GWs}
\label{subsec:stack}
Measuring a low-frequency gravitational wave requires a long observation time, at least of the order $T \sim 1/f_{\rm{GW}}$. Here we devise a strategy to integrate gravitational wave signals over the total mission duration of $\sim$10 years, which consists of stacking many independent tracking runs with a shorter duration, taken repeatedly in the entirety of the expected Uranus mission duration. {We assume that a single tracking station is allocated to perform the tracking runs. Then, the duration of each individual tracking run is limited as the spacecraft will fall below the horizon of any given tracking station within a fraction of a day, severing the Doppler link and preventing the continuous tracking of a coherent signal. To model this complication we consider a typical tracking run of a single station to have a maximal duration of approximately $8$ hours, after which the link may only be reinstated after a clock reset. This amounts to either switching tracking stations or waiting for the next day.}

Thus, the first step is to be able to generate noise realisations for single 8-hour tracking passes as a time series, while preserving the properties defined by the PSD. We construct the hypothetical noise time series $y_2^{\rm{noise}}$ of an entire 10 year tracking run by performing a discrete inverse Fourier transform of $S_n$, where we take the \textit{Cassini} noise spectral density as shown in Fig.~\ref{fig:cass_noise}. {Expressed as a Fourier series, the time domain noise reads:}
\begin{align}
\nonumber
    y_2^{\rm{noise}}(t) = 
     \sum_{f_i \in \left[f_{\rm{min}},f_{\rm{max}} \right]} 2\mathcal{A}_i\sqrt{f_{\rm{min}}S(f_i)} \cos( f_i t +\alpha_i) .
    \label{eq:noise_time_series}
\end{align}
Here $f_{\rm{min}} = (10 \, {\rm{yr}})^{-1}$ is the inverse of the cruise duration and $f_{\rm{max}} = \tau ^{-1}$ is the sampling rate. {The coefficients $\mathcal{A}_i$ together with the phases $\alpha_i$ determine a specific realisation of noise. In the limit of Gaussian noise, the phases are uniformly sampled from $[0, 2 \pi)$. The amplitudes $\mathcal{A}_i$ are instead drawn as random variables from a Gaussian with standard deviation of 1. The frequencies $f_i$ are given by all the integer multiples of $f_{\rm{min}}$ up to $f_{\rm max}$, as in a standard Fourier series.}

\begin{figure*}
    \centering
    \includegraphics[width = \textwidth]{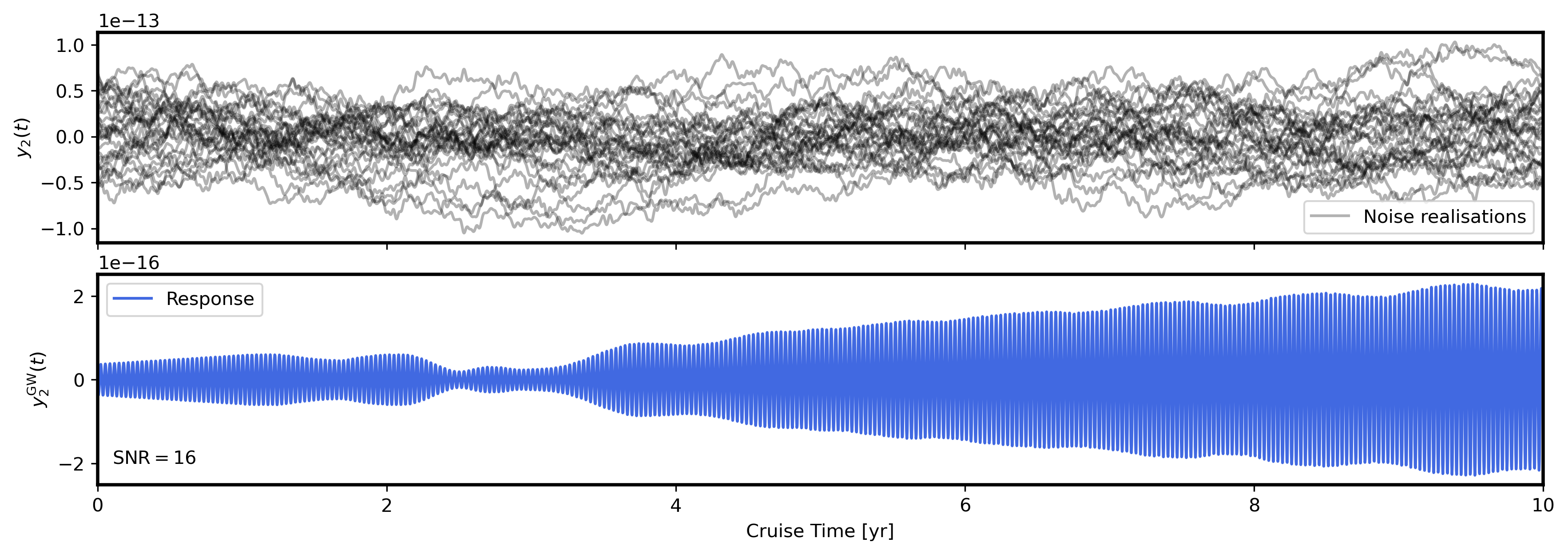}
    \caption{The two-way frequency fluctuation $y_{2}$ due to 30 different noise realisations (black lines), and due to the isolated GW signal (blue line) for 3650 different 8-hour measurements conducted over the spacecraft cruise. The change in the shape of the GW response is due to the variation of the spacecraft-Earth vector with respect to the GW source as the spacecraft cruises interplanetary space. {The quantities responsible for this variation are the light travel time $T_2$ and the projection $\mu$, which change non-trivially over the spacecraft trajectory (see also Sec. \ref{sec:senscurves}).
    The injected monochromatic GW has an amplitude of $10^{-15}$ and a frequency of $10^{-6}$ Hz, emanating from a face-on source 90$^{\circ}$ off of the ecliptic plane. This results in an SNR of approximately 16.}}
    \label{fig:track}
\end{figure*}

We can simulate {tracking data from a virtual mission by evaluating Eq.~\eqref{eq:noise_time_series} and adding it to the coherent Doppler response to a GW signal over the time segments during which the Doppler tracking is active}. To re-iterate, we model the mission's total tracking data as being composed of a sum of the GW response and the noise realisation, stacked over $N$ consecutive tracking runs:
\begin{align}
    y_2^{\rm{tot}} = \bigoplus_{n \leq N} \big( y_2^{\rm GW}(t_n;t) + y_2^{\rm noise}(t_n;t) \big),
\end{align}
where the tensor sum symbol describes the stacking of several, independent 8 h passes with arbitrary gaps between them. Here $t_n$ corresponds to the timestamp at which the $n$-th tracking run is initiated, and $0 < t< 8$~hours follows a clock that is started anew at the beginning of every tracking run. A possible realisation of the virtual tracking data, where the gaps have been removed, is visualised in Fig.~\ref{fig:track}. We note that from here on, we will assume that the tracking runs are spaced regularly over the mission duration (see Section \ref{sec:noise}). While this is most likely not realistic, investigating the distribution of tracking runs is only useful once an official mission trajectory is determined. To summarise, in addition to the Allan deviation, each realisation of $y_2^{\rm{tot}}$ is thus determined by the parameters characterising both the GW and the spacecraft trajectory. Schematically, they are given by the following:
\begin{align*}
    \text{GW } \to &\begin{cases}
        \hat{\boldsymbol{k}}; & \text{Propagation vector}\\
        f_{\scriptscriptstyle\rm GW}; & \text{Frequency}\\
        A; & \text{Amplitude}\\
        \phi_0; & \text{Phase}
    \end{cases}\\
    \text{Trajectory} \to &\begin{cases}
        \hat{\boldsymbol{n}}(t_n;t); & \text{Link unit vector}\\
        T_2(t_n;t); & \text{Two-way travel time}
    \end{cases}
\end{align*}
We note that, since the trajectory parameters $\left[\hat{\boldsymbol{n}}(t_n;t), T_2(t_n;t)\right]$ only evolve very slowly, we can essentially treat them as constant over a single 8-hour tracking run. Therefore, we only have to specify the mission's trajectory at the timestamps $t_n$:
\begin{align}
    \hat{\boldsymbol{n}}(t_n;t) &\to \hat{\boldsymbol{n}}(t_n), \\
    T_2(t_n;t) &\to T_2(t_n).
\end{align}
This simplification not only expedites the computations for upcoming sections, but also easily allows to efficiently repeat the analysis with different trajectories. This will become especially convenient once the official mission trajectory becomes available.

\section{GW sensitivity curves for the prospective Uranus Mission}
\label{sec:senscurves}
\subsection{Refresher on sensitivity curves}

In this section we estimate how well the GW signal can be extracted from the tracking noise {and define a general notion of the mission's sensitivity curve to GWs that can be coherently modelled, such as those emitted by massive black hole binaries. Typically, the sensitivity curve of a detector to a GW is defined from the detector
noise power spectral density $S_{\rm n}(f)$ and the transfer function $R(f)$, which encodes the response of the detector to an incoming GW signal with spectral density $S_{\rm GW}(f)$~\citep[\eg,][]{Larson:2000}. The spectral density of the full GW signal plus detector response, per GW cycle is, $S_{\rm GW}(f) R(f)$.
Then the optimal SNR of a GW signal is given by~\cite[\eg,][]{2019robson}:
\begin{align}
    \text{SNR}^2=4\int \frac{ 2f' T_{\rm obs}S_{\rm GW}(f') R(f')}{S_{\rm n}(f') f'}\, {\rm{d}}f'.
\end{align}
The sensitivity curve is then typically identified with the quantity:
\begin{align}
    \label{eq:typ_ses}
    A_{\rm sens}(f)= \sqrt{\frac{S_{\rm n}(f) f}{R(f)}},
\end{align}}
{where $T_{\rm obs}$ is the observation time and the response $R(f)$ is suitably averaged over the sky localisation and inclination of the source. Eq. \eqref{eq:typ_ses} is a convenient definition of sensitivity, because it allows to identify the area between the familiar GW characteristic strain $h_{\rm c} = \sqrt{2f T_{\rm obs}}\sqrt{S_{\rm GW}(f) f}$ and $A_{\rm GW}$ as the SNR:
\begin{align}
    {\rm{SNR}} = \sqrt{4 \times \int \frac{h_{\rm c}(f^\prime)^2}{A_{\rm{sens}}(f^\prime)^2} {\rm d} \ln f^\prime},
    \label{eq:SNRgw}
\end{align}
and for monochromatic sources SNR$\sim h_{\rm c}/A_{\rm sens}$.}

{Examples for the transfer function $R(f)$ for a fixed detector arm and incidence angle can be seen in Fig.~\ref{fig:cass_noise}, though typically the latter is averaged over the GW incidence angles~\citep{2006armstrong,2019robson,2024mcquinn}. However, in the particular case of a deep interplanetary ranging mission with a duration of 10 years, following the simple approach detailed above is not ideal. The reasons are the following:
\begin{itemize}
    \item Data is taken non-continuously over the mission duration, which affects both the spectral response and the definition of characteristic strain.
    \item The light travel time $T_2$ and the projection parameter $\mu$ vary significantly over the spacecraft trajectory. 
\end{itemize}
}
{The latter aspect is crucial, since the trajectory influences both the GW response and its PSD in a nontrivial manner. In particular, the presence of curvature in the trajectory significantly alters the projection of the wave onto the Doppler-link\footnote{While this complicates the calculation of the sensitivity curve, it is ultimately a welcome aspect as it may allow to better localise sources in the sky.}. The result is that the transfer function $R_{\rm GW}(f)$ cannot be easily separated from the pure GW PSD (see Ref.~\cite{Larson:2000} for the derivation of $R_{\rm GW}(f)$ for a fixed length and angle tracking system). The optimal SNR of the Doppler system is instead directly computed from:
\begin{align}
    {\rm{SNR}}^2 = 4\int\frac{2 f' T_{\rm obs} S_{\rm y_2}(f')}{S_{\rm n}(f')f'}\,{\rm{d}}f',
\end{align}
such that we do not have to separate $S_{\rm y_2}$ into a pure GW part and a source response part. While on average we expect the sensitivity to still be approximated by the function $A_{\rm sens}= \sqrt{fS_{\rm n}/R}$, this will not necessarily be representative depending on the specific UOP interplanetary orbit and the individual injected wave. Therefore, here we devise a more robust strategy to define a sensitivity curve, which can later be used for a more thorough investigations of individual signals and of the UOP planned trajectory.}

\subsection{Noise and response Lomb-Scargle PSDs}\label{sec:psds}
\begin{figure}
    \centering
    \includegraphics[width = \columnwidth]{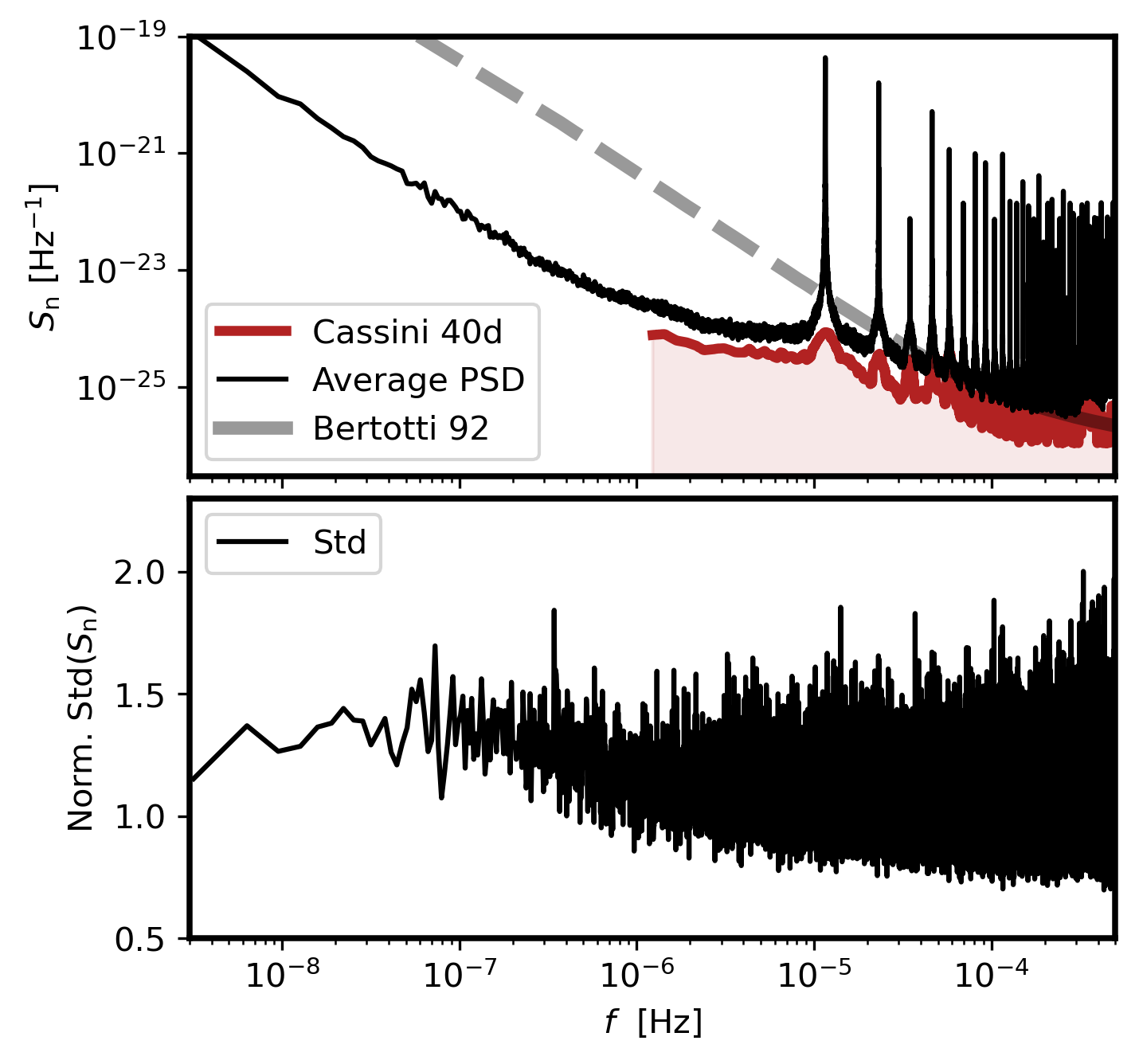}
    \caption{{Top panel: The Lomb-Scargle PSD of the noise in the Doppler tracking system, averaged over 100 realisations of Eq. \eqref{eq:noise_time_series}. Note the distinctive peaks in the PSD at $f= 1/16$~h$^{-1} \sim 1.7 \times 10^{-5}$~Hz and multiples thereof, caused by the uneven sampling. Bottom panel: The standard deviation of the noise PSD realisations at every frequency bin, for 100 realisations, normalised with the noise PSD value.}}
    \label{fig:avrgpsd}
\end{figure}

\begin{figure}
    \centering
    \includegraphics[width = \columnwidth]{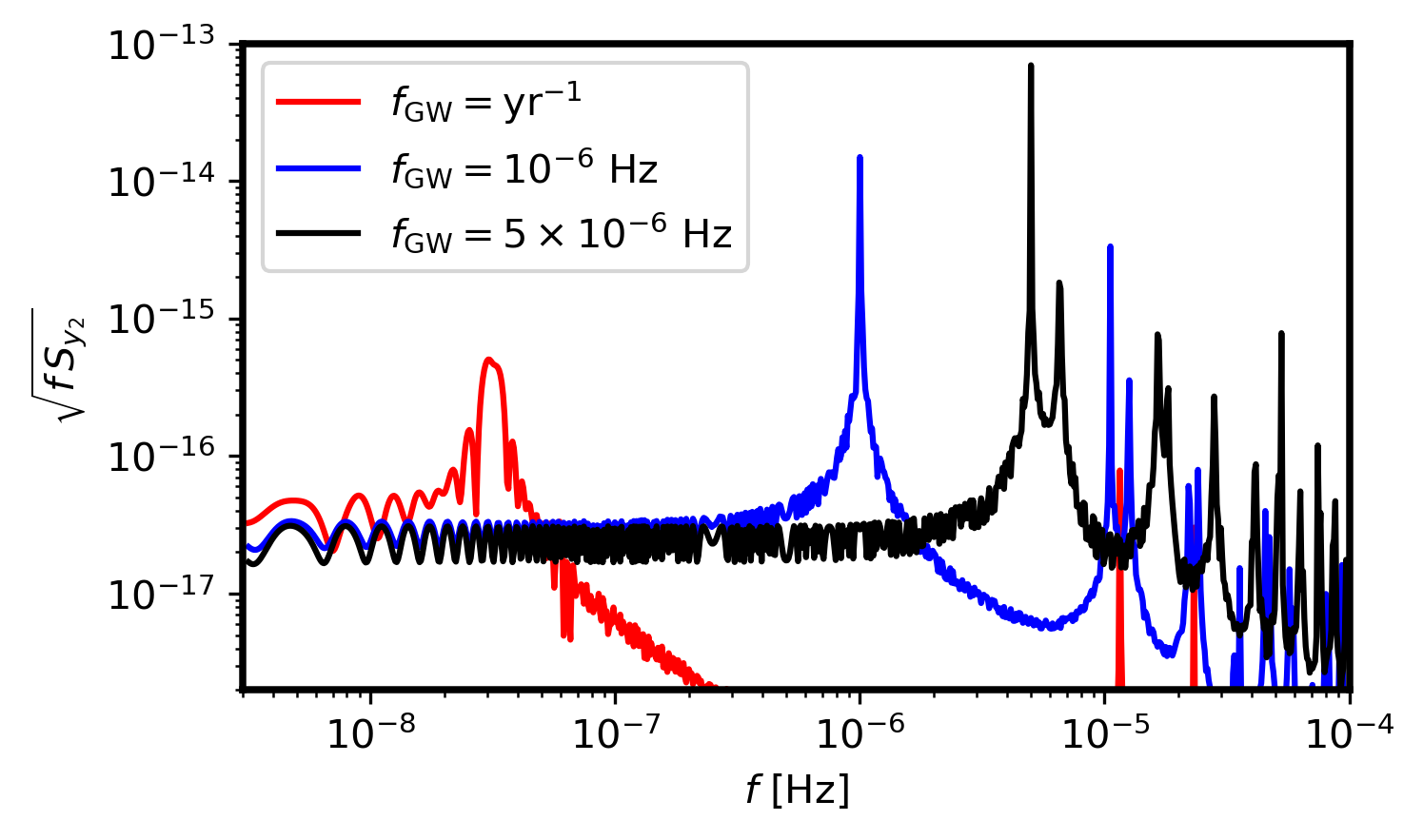}
    
    \caption{{Spectral responses of the UOP's Doppler tracking system to monochromatic GW of different frequencies and a fixed amplitude of $A = 10^{-15}$ and optimal initial orientation. Note how the response PSD deviates strongly from a pure delta function for GWs with $f\sim$yr$^{-1}$ or $f\sim 10^{-5}$ Hz. In the former case, the deviation is caused by the coupling between yr timescale variations in the Doppler tracking system's arm length and orientation. In the latter case the uneven sampling of the data introduces non trivial features in the PSD. }}
    \label{fig:response_psd}
\end{figure}

To define a sensitivity curve for the mission, one must first construct an appropriate noise power spectrum that accounts for the gaps in the tracking data. Our strategy is as follows:
\begin{itemize}
    \item Generate several realisations of the noise time-series according to Eq. (\ref{eq:noise_time_series}), evaluated on consecutive 8 hr tracking runs over a 10 yr period.
    \item Compute a number of Lomb-Scargle PSDs of the noise realisations.
    \item Compute the expected-distribution (average) of the PSD realizations to define a typical noise realisation.
    \item Calculate the standard deviation of the noise PSD distribution to define a detection statistic.
\end{itemize}
{Our baseline noise PSD is constructed from averaging 100 noise realisations. The noise is evaluated over 3650, 8 hr tracking runs with a resolution of 100 s, spaced evenly over the 10 yr cruise phase. The resulting average PSD and its standard deviation is plotted in Figure \ref{fig:avrgpsd}.}
{The second step is to compute the Lomb Scargle PSDs of the tracking system's response to monochromatic waves with various frequencies, amplitudes, phases and orientation angles. Figure \ref{fig:response_psd} shows three examples of response PSDs for GW of different frequencies. The time variation in the Doppler tracking system arm length and orientation, as well as the uneven sampling in the tracking runs, produce non-trivial deviations from a delta-function-like response.
To define a typical signal, we average the response PSD over the binary inclination angle $\iota$ and the sky location of the source with respect to the ecliptic plane. We highlight here that the response of individual signals will greatly vary from the averaged case. Interestingly, the mission lacks "blindspots" in for any longitudinal angle due to the presence of curvature in the UOP trajetcory ({See Appendix \ref{app2} for a plot of the expected SNR as a function of source angles). Otherwise, the averaging procedure is analogous to standard practive in GW detectors \citep[see e.g.][]{2019robson}}}

\subsection{{Defining the sensitivity to monochromatic GWs}}
\begin{figure}
    \centering
    \includegraphics[width = \columnwidth]{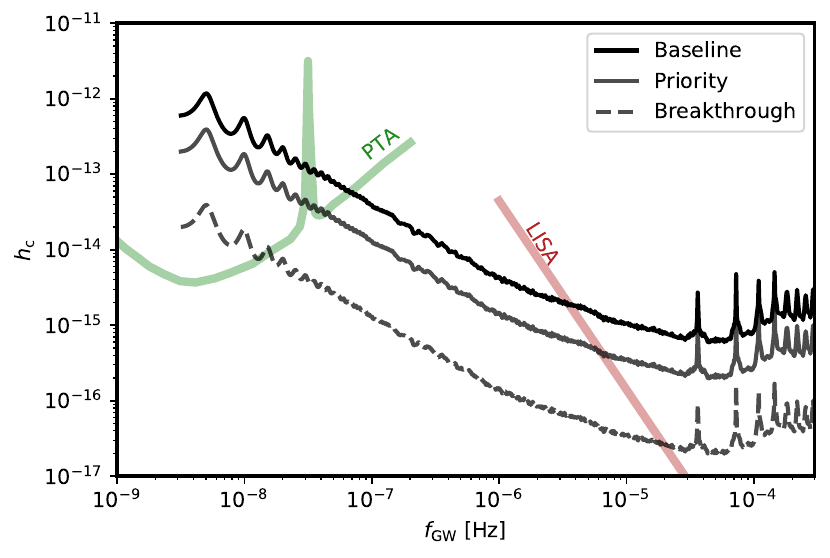}
    \includegraphics[width = 1\columnwidth]{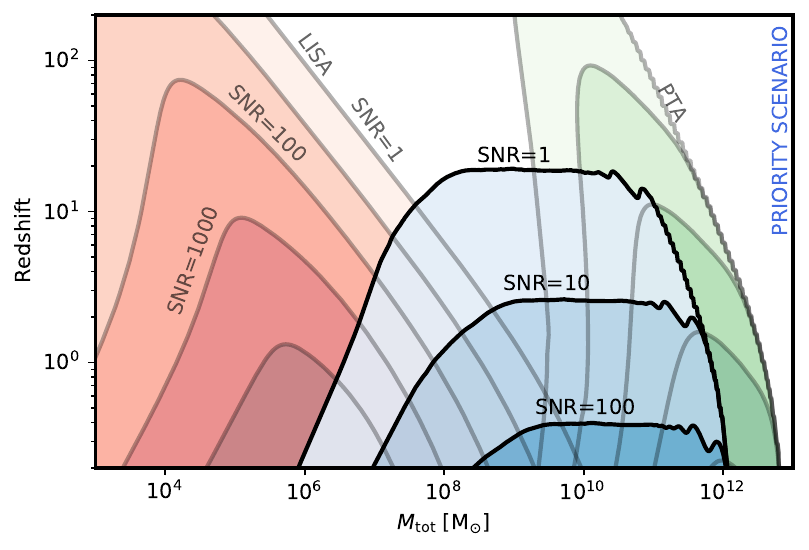}
    \caption{In the top panel, we compare the GW sensitivity of the prospective Uranus mission to the current sensitivity of PTAs~\citep{2023epta} and the sensitivity of the recently adopted LISA mission. The three black/grey lines correspond to the scenarios detailed in Section \ref{sec:senscurves}. In the bottom panel, we show a waterfall plot of the mission's hypothetical detection horizon for \textit{chirping} equal-mass (here masses are in the observer's frame) massive black hole binaries, assuming a maximum of 10 years for the signal duration and the \textit{priority} scenario, as described in Section \ref{sec:scenarios}. {We compare it to the same plot for PTAs and for LISA (assuming 4 years of maximal signal duration). The SNR iso-lines start at SNR=1 and are denote factor 10 increases in SNR.} Note that monochromatic sources, expected for PTAs and from our population models in \S \ref{S:smbh}, will have lower SNR for the same mass and redshift in the bottom panel.
    }
\label{fig:Sens_water}
\end{figure}

{Figure \ref{fig:response_psd} shows how the response of the UOP Doppler tracking system deviates from a simple delta function due to the variation in the system's arm length and orientation, as well as the uneven sampling. Typically, the usual definition of a sensitivity curve $A_{\rm sens}= \sqrt{S_{\rm n}/R}$ is useful since for monochromatic GW, SNR$\approx h_{\rm c}/A_{\rm sens}$. This is no longer true as soon as the response deviates significantly from a delta function. For a time-varying Doppler tracking system, it is more useful to use a slightly modified definition of sensitivity, which already entails the integration over the frequency span of the signal PSD. Recall that the Doppler response of a Doppler tracking system (Eq.~\eqref{eq:y2}) scales with the amplitude of the monochromatic wave as $A$. Therefore, $S_{\rm y_2}^{\rm GW} \propto A^2$ and:
\begin{align}
    {\rm{SNR}}^2 = 4\int\frac{2 f' T_{\rm obs} S_{\rm y_2}^{\rm GW}}{S_{\rm n}} \, {\rm{d}} \log f' \propto A^2.
\end{align}
To define a sensitivity curve, we evaluate the integral numerically for monochromatic GW of various frequencies $f$ and subsequently solve for the amplitude $A$, such that SNR$=1$. As a final step, we factor out the number of cycles by reducing the sensitivity with a factor $\sqrt{2fT_{\rm obs}}$. Then we are able to identify the full solution as a function of frequency $A_{\rm sens}^{{\rm{ SNR}}=1}(f)$ as the mission's sensitivity to GWs. For a monochromatic signal with amplitude $A$, the SNR is then simply given by $\sqrt{2fT_{\rm obs}}(A/A_{\rm sens}^{{\rm{ SNR}}=1}(f))$, in exact analogy to usual GW wave sensitivity curves. Note that here we must introduce the concept of the mission's tracking duty cycle $d_{\rm c}$, where:
\begin{align}
  d_{\rm c} = \frac{{\rm{total \, \, tracking \,\, time}}}{\rm{mission \, \, duration}},
\end{align}
for which we assume a typical value of $d_{\rm c}= 1/3$ to represent a single 8 hr tracking run per day. The observation time $T_{\rm obs}$ is therefore replaced by $d_{\rm c} \times 10$ yr. Note that in this paper we do not consider the fact that higher frequency signals are more likely to originate from chirping sources, whose observation time may be capped by the binary inspiral timescale $T_{\rm in}$. While some of this effect may be reabsorbed by replacing $T_{\rm obs}$ with $T_{\rm in}$ in the SNR computation, the exact timing of the chirp and merger event along the mission duration do indeed affect the response in a way that has to be quantified numerically. We leave a more thorough analysis of chirping signals to future work. Within these assumptions, the sensitivity curve of the mission is unique. We compare it with other existing and proposed GW observatories in Fig.~\ref{fig:Sens_water}.}

{
To conclude the section, we discuss aspects related to detection statistics, in order to quantify how reliably a signal with a given SNR can be distinguished from noise. Consider that a detection will consist of a total signal $s(t)$ composed of the waveform $h(t)$ and a noise realisation $n(t)$. For Gaussian noise likelihood ratio $R_{\rm L}$ to distinguish $s(t)$ from a pure noise realisation is:
\begin{align}
    R_{\rm L} = e^{(s,h)}e^{-(h,h)/2}
\end{align}
where we used the standard definition of the noise averaged inner product \cite{Maggiore:2007ulw,2011creighton}:
\begin{align}
    (a,b) = 2\int_0^{\infty}\frac{\tilde{a}\tilde{b}^* + \tilde{a}^*\tilde{b} }{S_{\rm n}}{\rm d}f.
\end{align}
Given that $s = h+ n$ we have:
\begin{align}
    R_{\rm L} =e^{(h + n,h)}e^{-(h,h)/2}= e^{(n,h)} e^{(h,h)/2}.
\end{align}
For the purposes of obtaining a concrete result, we make the simplification that noise is completely uncorrelated with the signal, i.e $(n,h)\sim 0$. Then we have:
\begin{align}
    \log R_{\rm L} \approx (h,h)/2 = {\rm SNR}(h)^2/2.
\end{align}
The required odds ratio is determined by a number of statistical considerations, including for example the look-elsewhere effect. In this case , the threshold for detectability depends on the number of templates used to search the data. For LIGO/Virgo/KAGRA, the quoted value translates to a threshold SNR of typically SNR=8. In this work, we use the values of SNR=1, 3 and 8 as references without presenting a thorough calculation of the required likelihood ratios. However, we expect the relevant threshold here to be significantly lower than for LVK, since long lived monochromatic signals require templates with fewer parameters and are present for the entire duration of the data. Finally, we note that a related option to define a detection statistic is the so called power statistic \cite{Maggiore:2007ulw} which estimates the probability of a random noise realisation to have as much excess power over the expected noise PSD as the putative signal. As seen in Fig. \ref{fig:avrgpsd}, our average noise PSD has a constant typical
standard deviation of approximately $\sim1.5\times S_{\rm n}$ over a
wide range of frequencies. Therefore, a random noise realisation will have excess power of a factor 1.5 (3) in
approximately 16\% (2.3\% ) of the realisations, assuming a Gaussian distribution as an approximant of the
$\chi^2$ distribution. Note that the distribution in power is exactly related to the distribution in SNR$^2$.}

\subsection{{Defining the sensitivity to stochastic GWs}}

{Previous works, in particular Ref.~\citep{2003ApJ...599..806A}, have derived SGWB constraints from Doppler-tracking data by simply dividing the sky- and polarization-averaged power spectrum $S_{y_2}(f)$ by the (known) transfer function $\bar{R}_2(f)$ to obtain an upper limit on the GW power spectrum.
Here the situation is complicated by the evolution of the light-travel time $T_2$ and the presence of gaps in the tracking data, which modify the effective GW power spectrum in the data stream in much the same way that they affect the noise power spectrum.
We therefore account for these complications using the same approach described in Sec.~\ref{sec:psds} to model the effective GW power spectrum corresponding to a particular model of the SGWB spectrum, $\Omega_{\scriptscriptstyle \rm GW}(f)$, including changes in the transfer function due to the varying light-travel time.
The signal-to-noise ratio for this spectrum is then simply computed as~\cite{Maggiore:2007ulw}
\begin{equation}
    \mathrm{SNR}^2=\sum_i\frac{S_{y_2}^{\scriptscriptstyle \rm GW}(f_i)}{S_n(f_i)},
\end{equation}
with the sum running over frequency bins.}

\begin{figure}
    \centering
    \includegraphics[width = \columnwidth]{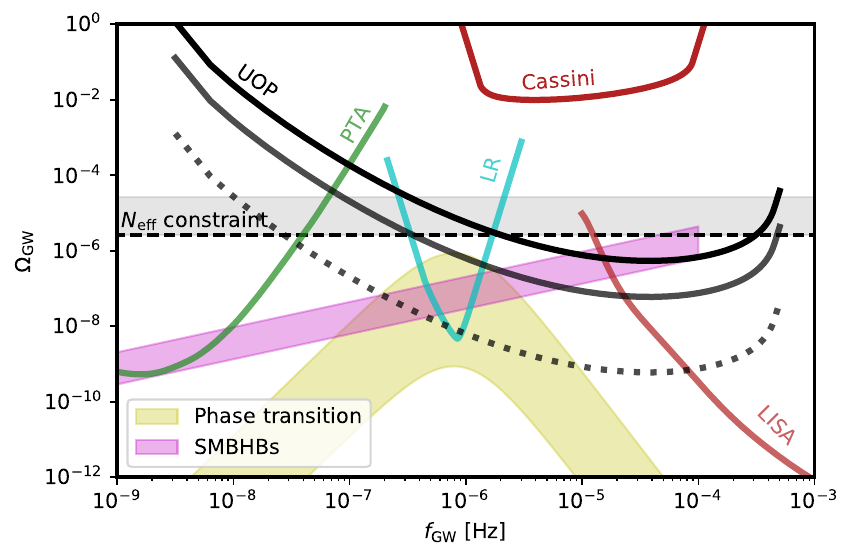}
    \caption{Forecast sensitivity to the SGWB spectrum $\Omega_\mathrm{GW}(f)$ for our baseline, priority, and breakthrough scenarios (solid and dotted black curves), along with existing sensitivities of the NANOGrav PTA (green)~\cite{NANOG-SMBHBs:2023} and Doppler tracking with \textit{Cassini} (red), as well as forecast sensitivities of LISA (brown)~\cite{2019robson} and binary resonance searches with lunar laser ranging data (LR, light blue)~\citep{Foster:2025nzf}.
    The black horizontal line ($N_{\rm eff}$ constraint) shows the indirect constraint on the \emph{integrated} SGWB intensity from measurements of the Hubble rate during radiation domination.
    The pink shaded region indicates the astrophysical SGWB from unresolved SMBHBs (with amplitude inferred from NANOGrav and fixed $\propto f^{2/3}$ spectral tilt, as appropriate in the absence of environmental or relativistic effects).
    The yellow shaded region indicates a potential cosmological signal from a first-order phase transition in the early Universe.
    }
    \label{fig:omega_gw}
\end{figure}

{For a SGWB of a fixed spectral shape, this process allows us to derive an upper limit on the overall amplitude of the signal, corresponding to $\mathrm{SNR}=1$.
By computing the upper limit for a family of power-law spectra $\Omega_{\scriptscriptstyle \rm GW}\propto f^\alpha$ (with $\alpha\in[-10,10]$), we construct a `power-law integrated' (PI) curve~\cite{Thrane:2013oya}, shown in Fig.~\ref{fig:omega_gw}.
This curve has the property that any SGWB signal that intersects it will have $\mathrm{SNR}\ge1$.}

{Note that unlike the monochromatic signals discussed above, it is not possible to carry out a coherent matched-filter search for the SGWB.
The procedure described above is instead an incoherent ``excess power'' search, which relies crucially on knowing the noise power spectral density of the mission, $S_n(f)$.
This highlights the importance of characterizing and calibrating the various contributions to this PSD for the UOP mission.
We note that it may be possible to leverage prior information about potential SGWB signals of interest---in particular, the amplitude and spectral shape of the SGWB from unresolved SMBHBs are already constrained by PTA data and theoretical modelling of this population.}

\subsection{Baseline, Priority and Breakthrough scenarios}
\label{sec:scenarios}
{
The monochromatic sensitivity curves constructed above (Fig.~\ref{fig:Sens_water}) span the interesting frequency range of few $\sim 10^{-9}$ Hz to $\sim 10^{-4}$ Hz. Below $\sim 10^{-6}$ Hz, the slope is approximately $f^{-1}$, where the sensitivity is determined by the extrapolation of the \textit{Cassini} noise.
%
At higher frequencies, the slope mellows to approximately $\sim f^{-1/2}$ before a turning point at $f\sim$1/(8 hours) where sensitivity is additionally affected by the discretisation of the observation period. We truncate it at few $\times 10^{-4}$ Hz to assure that numerical noise due to the Nyquist frequency is completely absent. The features of the sensitivity are a result of the interaction between the GW response and the time changing detector geometry, the un-even sampling of the tracking data as well as inheriting the noisiness of the observed \textit{Cassini} PSD. Crucially, strain sensitivities of the order few $\sim 10^{-14}$ in the micro-Hz band can be achieved without requiring significant improvements in the Allan deviation, {\textit{provided that the mission's duty cycle is of order $\sqrt{d_{\rm c}}\sim 1$, \ie, that tracking data is accumulated over a substantial amount of the entire mission duration}}.
We note that this amount of tracking is not typical for deep interplanetary missions. Here, however, we argue that it should be prioritised in parallel with Allan deviation improvements in order to achieve the scientific objectives detailed in this manuscript. We assume this level of commitment to the tracking and fix a duty cycle of $d_{\rm c} = 1/3$, which arises from a total of approximately 3650, 8-hour tracking runs. We note here that switching to multiple tracking stations to achieve $d_{\rm c} > 1/3$ should not, in principle, introduce additional sources of noise as long as the phase of the GW is not lost due to a lack of clock synchronization between the stations. Considering the low frequency of the tracked GWs, this is not an issue here.  \newline \newline We then define three possible future scenarios, in which different technological improvements are implemented with respect to Cassini era ranging. 
\newline \newline
In the {\textit{baseline}} scenario, we do not assume any improvement with respect to \textit{Cassini}-era tracking technology. The Allan deviation of this scenario is then $\sigma_{\rm A}(10\, \mathrm{s})= 10^{-14}$, at a sampling rate of 10 seconds. In this scenario, moderate sensitivity gains with respect to $\textit{Cassini}$ are still expected due to the longer detector arm, as well as the absence of blind spots due to the changing detector geometry along the cruise.
\newline \newline
In the {\textit{priority}} scenario, we assume that optimising currently available tracking technology becomes one of the explicit targets of the mission before launch. We assume that the Allan deviation at 10 seconds is reduced to the optimal \textit{Cassini}-era value of $3\times 10^{-15}$, which was achieved with a longer averaging time of $\tau = 10^4$ s. 
\newline \newline
In the {\textit{optimistic}} scenario, we assume that technological improvements in tracking technology (see, \eg, Sec.~\ref{sec:noise}) will have acquired an appropriate readiness level for an interplanetary mission. As a proxy for such technological improvements, we assume an Allan deviation at 10 seconds of $10^{-15}$. We discuss the prospects to achieve such low levels of noise in Section \ref{subsec:optical}. Here we simply note that by the time the UOP is operational, almost half a century of technological improvement will have taken place since the launch of the \textit{Cassini} mission.}
\newline \newline

\section{Detection forecast for Black Hole binaries}
\label{S:smbh}

Supermassive black hole binaries (SMBHBs) are among the loudest sources in the low-frequency GW sky. They are expected as a consequence of galaxy mergers, although how frequently they form and how they evolve in their environments is still not fully understood~\citep[see, \eg, the seminal papers][]{Begel:Blan:Rees:1980, 2003volontery,2005volker,2005hopkins,2013kormendy}. Recent constraints on a GW background signal by pulsar timing arrays (PTA) provide evidence of an SMBHB population in the mass range $\sim 10^8-10^{10} M_{\odot}$~\citep{NANOG-GWB:2023, EPTA-GWB:2023, ParkesPTA-GWB:2023, CPTA-GWB:2023} while the direct detections of lighter SMBHB mergers below a few $ 10^7 M_{\odot}$ is expected in the 2030s with the advent of space-based interferometers such as LISA~\citep{redbook} and TianQin~\citep{2016tian}. As we will show, the sensitivity derived in Section \ref{sec:senscurves} facilitates the detection of SMBHBs that are evolving beyond the nano-Hz band and into the milli-Hz regime. As an illustration, we show the hypothetical sensitivity of the mission to chirping, equal-mass BH binaries as a waterfall plot in the bottom panel of Fig.~\ref{fig:Sens_water}. The mission's peak sensitivity lies in the total mass range of $10^7$ to $10^{10} M_{\odot}$. Such heavy, chirping binaries would in principle be detectable with high confidence up to redshifts of $z\sim 3$ for an Allan deviation of $\sigma_{\rm A}=10^{-14}$ and essentially over the whole cosmological volume for $\sigma_{\rm A}=1\times 10^{-15}$.

However, a detection forecast requires us to consider the actual population of sources we expect to fall in this frequency band, and whether the majority will be detectable as a monochromatic signal rather than a chirp. Here we adopt two separate approaches to estimate the redshift-dependent merger rate of such sources.

\subsection{Model 1: Millennium simulation}
Here we detail a simple prescription that links the massive black hole merger rate to the much more established halo merger rate, based on the two Millenium simulations~\citep{2010fakhouri}. The latter works provide a convenient fit characterising the differential halo merger rate:
\begin{align}
    \label{eq:millennium}
    \frac{{\rm d}^2 \Gamma}{{\rm d}\xi {\rm d}z} = B_1 \left( \frac{M_{\rm{halo}}}{10^{12} \,{M}_{\odot}} \right)^{b_1} \xi ^{b_2} \exp \left[\left( \frac{\xi}{B_2} \right)^{b_3}\right] (1+z)^{b_4},
\end{align}
where $\Gamma$ is the total number of mergers that a halo of mass $M_{\rm{halo}}$ experiences over cosmic time, $\xi\leq1$ is the halo merger mass ratio and the best-fit parameters are given by $(B_1,B_2,b_1,b_2,b_3,b_4) = (0.0104,9.72 \times 9.72,0.133,-1.995,0.263,0.0993)$. We link Eq.~\eqref{eq:millennium} to the SMBH merger rate $\dot{N}_{\bullet \bullet}$ by multiplying the halo merger rate with the black hole mass function:
\begin{align}
    \label{eq:generalSMBBHrate}
    \frac{{\rm d} ^{3} \dot{N}_{\bullet \bullet}}{{\rm d}M_{\bullet} {\rm d}\xi {\rm d}z} &= {P_{\rm occ}}(M_{\rm{halo}},z) \frac{4 \pi c D^2_{\rm{com}}(z)}{(1+z)^3} \nonumber \\ &\times\frac{{\rm d}n_{\bullet}}{{\rm d}M_{\bullet}} \frac{{\rm d}^2 \Gamma}{{\rm d}\xi {\rm d}z}(\xi,z_{\rm{del}}),
\end{align}
where $D_{\rm{com}}$ is the comoving distance at redshift $z$ and we must additionally supply an occupation fraction $P_{\rm occ}$ of black holes in halos and a delay prescription $z_{\rm{del}}$ between the nominal halo merger time and the actual black hole merger time. Here we use the SMBH mass function ${{\rm d}n_{\bullet}}/{{\rm d}M_{\bullet}}$ as reported in Ref.~\cite{2013shankar} and adopt a simple relation between the halo and SMBH mass from Ref.~\cite{2009croton}:
\begin{align}
M_{\bullet}&=\left[\frac{M_{\rm{halo}}(1+z)}{2\times10^7\, M_\odot}\right]^{3/2}\, M_\odot,\label{eq:M_bullet}
\end{align}
which is consistent with both simulations and observations of massive galaxies~\citep{2021Marasco,2023bansal}.

We can now establish the number of detectable sources of gravitational radiation by integrating the differential SMBH merger rate. We define an SNR threshold $C$ and count all SMBH binaries that exceed it: 
\begin{align}
\label{Eq:Ndet}
   N_{\bullet \bullet}^{\rm{det}} = \int \int \int \int \frac{{\rm d}^{3} \dot{N}_{\bullet \bullet}}{{\rm d} M_{\bullet} {\rm d} \xi {\rm d} z} \frac{1}{\dot{f}_{\scriptscriptstyle\rm{GW}}} \Theta \left({\rm{SNR}-C}\right) \\ \nonumber
    \times{\rm d} f_{\scriptscriptstyle\rm{GW}} \,{\rm d} \xi {\rm d} M_{\bullet} {\rm d} z,
\end{align}
where $\Theta$ is a Heaviside function, and we distribute all binaries in frequency bins according to their residence times $\dot{f}_{\scriptscriptstyle\rm{GW}}^{-1}$ {where $f_{\rm GW}$ is the GW frequency in the observer's frame}. It is necessary to make two additional simplifications.
Firstly, we model the occupation fraction $P_{\rm occ}$ as an unspecified constant. For massive black holes $M > 10^{7} M_{\odot}$, simulations~\citep{2018ricarte} and observations of AGN~\citep{2019MNRAS.487..275G} show that the occupation fraction $P_{\rm occ}$ is a fraction of order unity. Secondly, we also neglect the effect of time delays between the halo merger and the BH merger, as they typically only amount to $10^8$--$10^9\,\rm years$ for the mass range we are considering~\citep{2015ApJ...810...49V,2017ApJ...840...31D}. This simplification is justified a posteriori, given that tracking will only ever be sensitive to binaries at $z \lesssim 1$ for realistic populations,\footnote{The seeming discrepancy with the detection horizon in Fig.~\ref{fig:Sens_water} derives from the detectability of monochromatic sources vs. the hypothetical chirping sources drawn for consistent comparison between bands in Fig.~\ref{fig:Sens_water}.} where a delay of 10$^9$ years at most corresponds to a small redshift shift $\Delta z_{\rm{del}} \sim 0.3$. 


\begin{figure}
    \centering
    \includegraphics[width=\columnwidth]{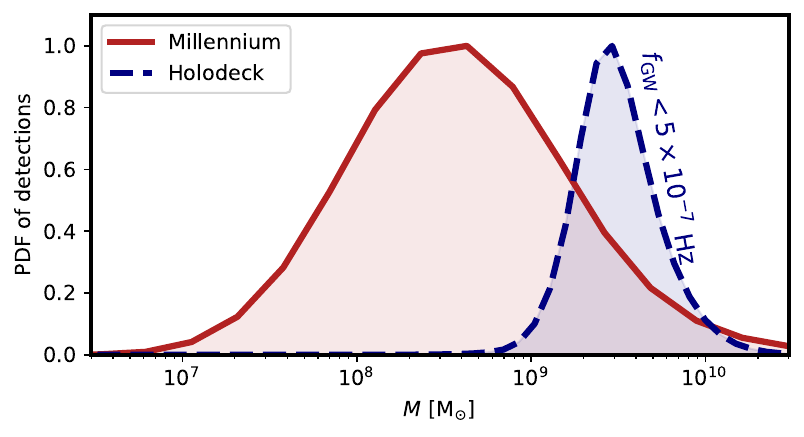}
    \includegraphics[width=\columnwidth]{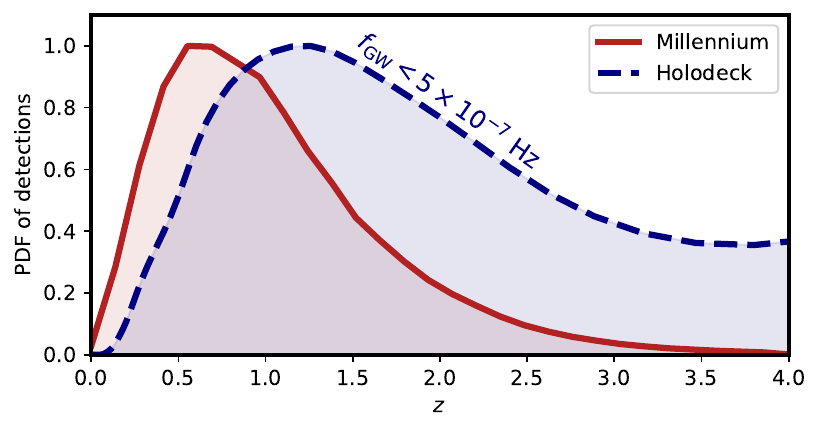}

    \includegraphics[width=\columnwidth]{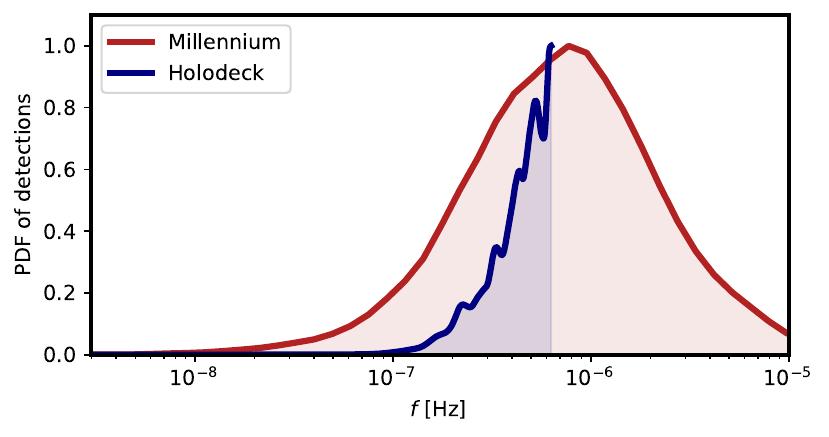}
    \caption{The differential distribution of detections 
     for both the Millennium and \holodeck{} (here chosen to be the median percentile realisation) population models, normalised with their respective maximum values. The top panel shows the distribution in mass decades $dN/d\log_{10}(M)$, while the middle panel shows the distribution in redshift $dN/d\log_{10}(z)$. The curves are displayed for the optimistic baseline with an SNR threshold of 1, but apply with little variation for all scenarios and SNR thresholds. The lowest panel shows the distribution of sources in $dN/d\log_{10}(f_{\rm GW})$. The results for \holodeck{} are truncated at high frequency where the model is not reliable.}
    \label{fig:populationdistr}
\end{figure}

\begin{table*}[]
\hspace{0.4cm} {Baseline; All sources} \hspace{2.5 cm} {Priority; All sources} \hspace{2.0cm} {Breakthrough; All sources} \newline
\noindent\rule{16cm}{0.4pt}\newline \newline
    \vspace{0.3cm}
    \centering
    \begin{tabular}{c|c|c}
       Pop.  & SNR &Nr. Det \\
       \hline
        Millennium & {1} & 0.2$\times P_{\rm occ}$ \\
        Millennium & 3 & $0.02\times P_{\rm occ}$ \\
        Millennium & 8 & 0\\
        
        
    \end{tabular}
    \hspace{1 cm}
     \begin{tabular}{c|c|c}
       Pop.  & SNR &Nr. Det \\
       \hline
        Millennium & 1 & 1.2$\times P_{\rm occ}$\\
        Millennium & 3 & 0.2$\times P_{\rm occ}$\\
        Millennium & 8 & 0.03$\times P_{\rm occ}$\\
        
    \end{tabular}
    \hspace{1 cm}
    \begin{tabular}{c|c|c}
       Pop.  & SNR &Nr. Det \\
       \hline
        Millennium & 1 & 24.2$\times P_{\rm occ}$\\
        Millennium & 3 & 5.9$\times P_{\rm occ}$\\
        Millennium & 8 & 1.6$\times P_{\rm occ}$\\

    \end{tabular} \newline
    
\vspace{0.5cm} \hspace{0.7 cm} 
{Baseline; $f < 5 \times10^{-7}$ Hz} \hspace{1.2 cm} {Priority; $f < 5 \times10^{-7}$ Hz} \hspace{1.2cm} {Breakthrough; $f < 5 \times10^{-7}$ Hz}  \newline
\noindent\rule{16cm}{0.4pt}\newline \newline
\vspace{0.3cm} 
  \hspace{-0.8cm} \begin{tabular}{c|c|c}
       Pop.  & SNR &Nr. Det \\
       \hline
        Millennium & {1} & 0.08$\times P_{\rm occ}$ \\
        Millennium & 3 & $0.009\times P_{\rm occ}$ \\
        Millennium & 8 & 0\\
        \hline
        \holodeck & 1 & $0.02^{+0.01}_{-0.002}$ \\
        \holodeck & 3 & $0$ \\
        \holodeck & 8 & $0$ \\
    \end{tabular}
    \hspace{0.7 cm}
     \begin{tabular}{c|c|c}
       Pop.  & SNR &Nr. Det \\
       \hline
        Millennium & 1 & 0.5$\times P_{\rm occ}$\\
        Millennium & 3 & 0.08$\times P_{\rm occ}$\\
        Millennium & 8 & 0.01$\times P_{\rm occ}$\\
        \hline
        \holodeck & 1 & $0.03^{+0.1}_{0.02}$ \\
        \holodeck & 3 & $0$ \\
        \holodeck & 8 & $0$ \\
    \end{tabular}
    \hspace{1 cm}
    \begin{tabular}{c|c|c}
       Pop.  & SNR &Nr. Det \\
       \hline
        Millennium & 1 & 14.6$\times P_{\rm occ}$\\
        Millennium & 3 & 3.1$\times P_{\rm occ}$\\
        Millennium & 8 & 0.7$\times P_{\rm occ}$\\
        \hline
        \holodeck & 1 & $0.8^{+2.9}_{-0.7}$\\
        \holodeck & 3 & $0.11^{+0.4}_{-0.1}$ \\
        \holodeck & 8 & $0.015^{+0.04}_{-0.011}$ \\

    \end{tabular}
    %
    %
    \caption{Tables summarising the expected number of detections of SMBHBs for the UOP scenarios detailed in Section \ref{sec:scenarios}, over the entire mission duration. The detection estimates are performed with the two population models explained in Section \ref{S:smbh} and reported for three different SNR thresholds (see Fig.~\ref{fig:mcmc1} for a visualisation of the typical parameter posteriors given these SNRs assuming white noise). The Millennium results scale with the occupation fraction of SMBHs in massive galaxies, $P_{\rm occ}$. The \holodeck{} numbers are quoted as median $\pm$ $25\%$ and $75\%$ quantile values. {The lower tables only contain monochromatic sources up to a threshold frequency at which the holodeck results stop being reliable. Below the threshold frequency, the two populations are comparable if $P_{\rm occ}\sim 0.1$. The frequency distribution in \holodeck \,  suggests a higher fraction of chirping signals which may boost the numbers further.}} 
    \label{tab:popresults}
\end{table*}

\subsection{Model 2: Holodeck}
We present another estimate of the detectable SMBHB population via the \holodeck~code~\citep[][and see Ref.~\cite{NANOG-SMBHBs:2023}]{holodeck_paper}.
Importantly, we utilise \holodeck~to build population models that generate nano-Hz-frequency GWBs with an amplitude that is consistent with the GWB for which multiple PTA experiments have recently found strong evidence~\citep{NANOG-GWB:2023, EPTA-GWB:2023, ParkesPTA-GWB:2023, CPTA-GWB:2023}.

We utilise \holodeck's semi-analytical SMBHB population models. Specifically, these models combine an observationally constrained galaxy mass function from Ref.~\cite{Leja+202004}, with a galaxy merger rate from Ref.~\cite{Rodriguez-Gomez+201505} to generate galaxy mergers. An SMBH-mass versus bulge-mass relation from Ref.~\cite{2013kormendy} is then used to calculate SMBHB parameters, which are then self-consistently evolved from galaxy-scales to nano-Hz separations using a phenomenological binary evolution model developed for \holodeck{} that combines binary decay due to environmental interactions and GWs~\citep{NANOG-SMBHBs:2023}. The change in source redshift over the course of binary evolution is naturally included. Each of these model components depends on a number of uncertain parameters with measurements and uncertainties described in the cited studies.

To sample SMBHBs that are consistent with PTA observations, we generate 1080 \holodeck{} populations by sampling over the uncertainties in each model parameter\footnote{29 parameters are marginalised over in this way, sampling from the \texttt{PS$\_$Astro$\_$AStrong$\_$AAll-v1.0} parameter space.}. We then calculate the resulting PTA-measured GWB spectrum, fit a power-law spectrum to the lowest five frequency bins, and select the populations with amplitudes and spectral indices consistent with recent measurements.
In particular, we require the GWB amplitude (characteristic strain at a frequency of $1 \, \mathrm{yr}^{-1}$) that is $A_\mathrm{yr} \in [10^{-15}, 10^{-14}]$, and a power-law index (in power-spectral-density of timing residuals) that is $\gamma \in [-5.0, -2.5]$~\citep[see, \eg,][]{IPTA-2024XX}. This procedure is repeated for each UOP sensitivity scenario and threshold SNR, drawing independent populations for each combination. Out of the 1080 populations, $130 \pm 10$ pass the above selection cuts. Those populations are then integrated over with a cut in detection SNR similarly to Eq.~\eqref{Eq:Ndet} to determine detectable SMBHB systems.


\subsection{Detection forecast}
The results of our detection forecast are summarised in Table \ref{tab:popresults} and Fig.~\ref{fig:populationdistr}. Regarding the former, we compute the number of expected detections of individual binary SMBHs at three separate SNR thresholds (1, 3, 8), for the three mission scenarios of baseline, priority, and optimistic (\S \ref{sec:scenarios}). 
For the \holodeck{} results we quote the median, $25\%$, and $75\%$ quantiles derived from the $\sim130$ GWB-selected populations described above. For the Millennium results, we quote expectation values in terms of the occupation fraction of BHs in halos.
The expected number of detections is broadly consistent for the two adopted population models; the Millennium expectation values fall within the inner $50\%$ range of \holodeck{} values for $P_{\rm{occ}} \sim 0.1$.

{In the baseline scenario, which does not assume any technological improvement with respect to \textit{Cassini}-era technology, we would typically expect the detection of $\mathcal{O}(0)$ binaries. A mild improvement in the Allan deviation, \eg, to $3\times 10^{-15}$, boosts the number of detections to up to $\mathcal{O}(0.1)$ at very low SNR. The detection of a supermassive BH binary at SNR=3 is becomes likely for the breakthrough scenario. With some numerical experiments based on the Millennium simulation population estimates, we find that the overall number of detectable monochromatic signals $N_{\rm det}$ scales as:
\begin{align}
   N_{\rm det} \approx \mathcal{O}(1) \times P_{\rm occ} \left(\frac{3}{{\rm{SNR}}} \frac{ 10^{-15}}{\sigma_{\rm A}}\right)^{1.2},
\end{align}
\ie, approximately linearly with a combination of the desired SNR and the achieved Allan deviation of the mission. Overall, the results shown in Table \ref{tab:popresults} suggest that a factor 30 improvement in the Allan deviation over the baseline mission would guarantee the detection of a few well constrained signals from SMBH binaries, in a mass range that is yet to be explored. Indeed, as seen in Fig.~\ref{fig:populationdistr}, the majority of detected sources would lie around a mass of few$\times10^{8} M_{\odot}$, being detectable out to redshifts of $z\sim 1$. The inspiral timescale of the typical source in the Millennium population estimate, \ie, $M \sim 10^8$~M$_{\odot}$, mass ratio of $\sim 1$ and frequency $f\sim 10^{-6}$~Hz, is approximately 10~yr. Therefore, the typical signals in the UOP are at the transition between almost monochromatic sources and chirping. They appear at frequencies of hundreds of nHz to tens of $\mu$Hz, \ie, above the PTA band and below the LISA band (see bottom panel of Fig.~\ref{fig:populationdistr}). A more sensitive mission can pick up more monochromatic signals at lower frequencies, where a larger number of binaries reside.}

{The main difference between the Millennium and the \holodeck~population models lies in the frequency distribution of sources, as can be seen in the bottom panel of Fig.~\ref{fig:populationdistr}. The \holodeck \, model prefers slightly higher frequency sources and hints at a population of chirping binaries that may be detected at frequencies $> \text{few}\times  10^{-6}$ Hz. The apparent difference in mass and redshift distributions is also a consequence of the frequency cut applied to the GWB-constrained \holodeck~models; we chose not to extrapolate models constrained by PTA measurements at $\sim$nHz  beyond $5\times 10^{-7}$ Hz.
A comparison of the results of the two population models is given in Table \ref{tab:popresults}. Up to the same threshold frequency, \holodeck \, results are compatible with the Millennium results for $P_{\rm occ}\sim 0.1$.} As a final comment on the topic of SMBHBs, we briefly mention that tracking the UOP would provide stringent bounds on the existence of an intermediate-mass companion to the main SMBH of the Milky Way, \ie, SgrA*. Current constraints arise from a range of techniques, see, \eg, Ref.~\cite{Will:2023nlt,Naoz:2019sjx}. Quite relevant for this work are the models where the GWs emitted fall in the micro-Hz band. In Ref.~\cite{Naoz:2019sjx}, it was shown that several of these models generate GWs with strains at the Solar system of up to $10^{-12}$. From Fig.~\ref{fig:Sens_water}, it can be easily extrapolated that all hidden companions that would produce strains of order $\sim 10^{-13}$ over the frequency range of $10^{-8}$ Hz to $10^{-5}$ Hz would be detectable.

We show how the different mission scenarios perform in terms of constraining stochastic GW energy densities in Fig.~\ref{fig:omega_gw}. Already in the priority scenario, the UOP will be sensitive to the stochastic background of SMBH binaries over the frequency range of $10^{-6}$ Hz to $10^{-5}$ Hz, provided that the latter follows the scaling of $f^{-2/3}$ that is reported by, \eg, Ref,~\cite{NANOG-GWB:2023}. Once again, this fills the gap between PTA and LISA sensitivities.
\newline \newline

Combining all of these prospects, we have demonstrated that Doppler tracking with the UOP is uniquely suited to detect GWs in the frequency regime between PTAs and space-based interferometry, {and that Allan deviation improvements should be pursued and added to the mission's technological requirements}. Sources in this frequency range comprise heavy SMBH binaries of several $10^8 M_{\odot}$ approaching coalescence, inspiraling pre-LISA systems and intermediate-mass black holes in the Milky Way. Such detections would unveil a more complete picture of SMBH formation, growth, and interaction, complementing both existing and planned GW detectors in the low-frequency regime. In the event of a concurrent detection with PTAs or LISA, a UOP detection may also aid in the difficult task of localising SMBHB sources~\citep[\eg,][]{soyuer2021}.

\section{Forecast constraints on early Universe signals}
\label{sec:earlyuniverse}
As well as filling the gap in our sensitivity to SMBHBs, our proposed GW searches are also sensitive to potential cosmological signals from the early Universe.
There is a huge range of mechanisms for generating GWs at early times~\citep{Caprini18}, each of which, if detected, would give valuable new insights into fundamental physics.
In the context of Doppler tracking with the UOP however, the most interesting prospects are for signals that are peaked in a relatively small range of frequencies, which risk going undetected by other GW experiments if there is not sufficient coverage of the micro-Hz frequency band.

The quintessential example of such a peaked cosmological GW signal is that generated by a first-order phase transition (FOPT) in the early Universe.
These transitions, in which a fundamental quantum field escapes from a metastable state by nucleating `bubbles' of a new phase, are a generic prediction of many extensions to the Standard Model of particle physics~\citep{Caprini18,Caprini:2015zlo,Caprini:2019egz}, notably beyond the reach of even the most powerful particle accelerators on Earth. 
Collisions between nucleated bubbles, and the subsequent acoustic and turbulent motion of the thermal plasma, can produce strong SGWB signals that peak at a frequency set by the energy scale of the transition, with higher frequencies corresponding to higher energies and thus earlier cosmic epochs~\citep{Kamionkowski:1993fg}.
This peak frequency is commonly approximated as
    \begin{equation}
        f_*\approx19\,\mu\mathrm{Hz}\times\frac{k_\mathrm{B}T_*}{100\,\mathrm{GeV}}\frac{\beta/H_*}{v_w/c}\left(\frac{g_*}{106.75}\right)^{1/6},
    \end{equation}
    where $T_*$ is the temperature of the thermal plasma at the time when the GWs are generated, $\beta$ is the duration of the transition, measured in units of the Hubble rate $H_*$ at that epoch, $v_w$ is the terminal expansion velocity of the bubble walls, and $g_*$ is the number of relativistic degrees of freedom in the plasma, normalised here to the Standard Model value.
Focusing for simplicity on the contribution from sound waves in the plasma~\citep{Hindmarsh:2013xza}, which is dominant in many scenarios, the SGWB spectrum can be modelled as a broken power law,
    \begin{equation}
        \Omega_\mathrm{GW}(f)\approx\Omega_\mathrm{GW}(f_*)\times(f/f*)^3\left[\frac{7}{4+3(f/f_*)^2}\right]^{7/2},
    \end{equation}
    where the peak intensity is
    \begin{align}
    \begin{split}
        \Omega_\mathrm{GW}(f_*)\approx5.7\times10^{-6}&\times\frac{v_w/c}{\beta/H_*}\left(\frac{\kappa\alpha}{1+\alpha}\right)^2\left(\frac{g_*}{106.75}\right)^{-1/3}\\
        &\times\left[1-(1+2\tau_\mathrm{sw}H_*)^{-1/2}\right],
    \end{split}
    \end{align}
    with $\kappa$ an efficiency parameter and $\tau_\mathrm{sw}$ the lifetime of the sound-wave source, both of which can be modelled as functions of the parameters $\alpha$, $\beta$, and $v_w$~\citep{Caprini:2015zlo,Caprini:2019egz}.

\begin{figure}
    \centering
    \includegraphics[width = \columnwidth]{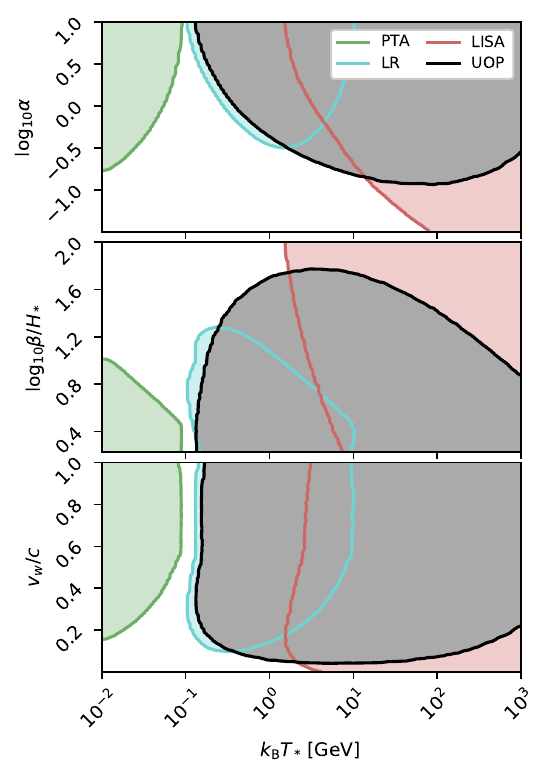}
    \caption{Forecast constraints on cosmological SGWB signals from early Universe phase transitions.
    These are shown using 2D projections of the 4D space of parameters describing the transition: the temperature of the Universe at the time of the transition, $T_*$, the fractional energy density released by the transition, $\alpha$, the rate of the transition, $\beta$, and the terminal velocity of the bubble walls, $v_w$.}
    \label{fig:fopt-constraints}
\end{figure}

Following Refs.~\citep{Blas:2021mpc,Blas:2021mqw}, we carry out a scan over the 4D parameter space $(T_*,\alpha,\beta,v_w)$ to assess the sensitivity of our proposed searches to FOPT signals.
As shown in Fig.~\ref{fig:fopt-constraints}, there is a broad swathe of parameter space for these models that is accessible with Doppler tracking searches in the micro-Hz band that would go undetected by pulsar timing arrays in the nano-Hz band and LISA in the milli-Hz band.
These constraints are complementary to those that have been forecasted for `binary resonance' searches for micro-Hz GWs using Lunar and satellite laser ranging data~\citep{Blas:2021mpc,Blas:2021mqw,Foster:2025csl,Foster:2025nzf}, allowing us to potentially detect or rule out a much larger family of FOPT models.

\section{Forecast constraints on ultralight dark matter}
\label{sec:uldm}

The expected constraints on $h_c$ from the top panel of Fig.~\ref{fig:Sens_water} already give an idea of the reach of UOP to detect ULDM (compare Eqs.~\eqref{eq:Doppler1} and~\eqref{eq:fshiftDM}). This suggests that for masses $m_{\scriptscriptstyle\rm DM}\gtrsim 10^{-22}$\,eV/$c^2$, UOP may represent the most sensitive probe for large regions in parameter space of dark matter models, even in the baseline scenario. Notably, this mass range is particularly relevant for ULDM phenomenology~\cite{Hui17,Ferreira:2020fam}. To illustrate this reach, let us consider the equivalent $h_c$ that would be produced by the three families of models we introduced in Section~\ref{sec:ULDMsignal}, and that could be summarised by Eqs.~\eqref{eq:psi_DM},~\eqref{eq:direct} and~\eqref{eq:shift}. Recall that $f_{\rm grav}\approx 2 f_{\scriptscriptstyle\rm DM}$, where $f_{\scriptscriptstyle\rm DM}$ can be found in Eq.~\eqref{eq:m_to_Hz} for the purely gravitational coupling and the quadratic direct coupling $\alpha(\phi)=\phi^2/\Lambda^2$. For the linear coupling, such as the one in Eq.~\eqref{eq:couplingaxion}, $f=f_{\rm DM}$.

\begin{figure}
    \centering
    \includegraphics[width =\columnwidth]{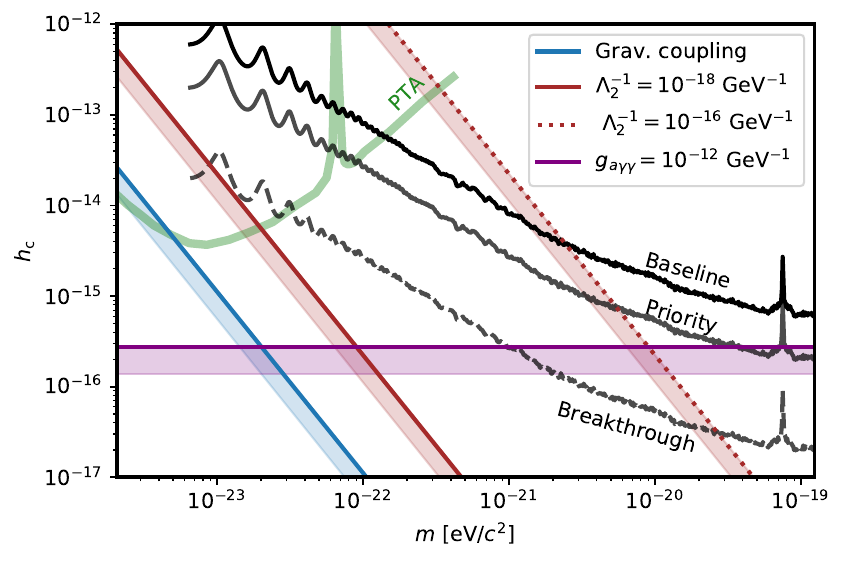}\\
    \includegraphics[width =\columnwidth]{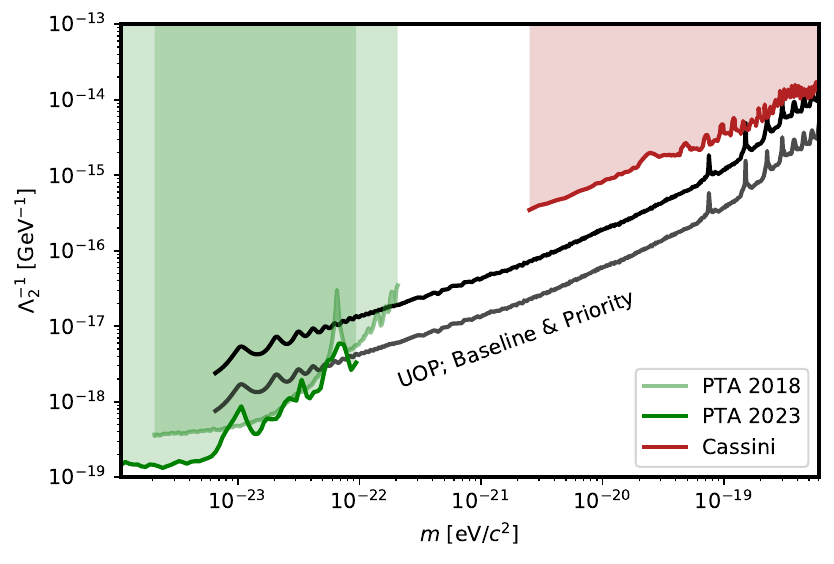}
    \caption{Upper plot: Expected sensitivity of the UOP for ULDM. The effect of ULDM is represented as a corresponding $h_c$ that can be directly compared with Fig.~\ref{fig:Sens_water}. We plot the gravitational case (Eq.~\eqref{eq:psi_DM}), two cases where ULDM couples to matter (Eq.~\eqref{eq:direct}), and a case where ULDM couples to light  (Eqs.~\eqref{eq:shift}). Lower plot: Expected sensitivity of UOP for ULDM for the model with direct coupling of matter to $\alpha(\phi)=\phi^2/\Lambda_2^2$. We compare with bounds from \textit{Cassini}~\citep{2003ApJ...599..806A} and the EPTA~\citep{EuropeanPulsarTimingArray:2023egv} (see also Ref.~\citep{Porayko:2018sfa}).}
    \label{fig:DM}
\end{figure}

The purely gravitational case represented by Eq.~\eqref{eq:psi_DM} (which is independent of other couplings) generates the black line in Fig.~\ref{fig:DM} as a function of the ULDM mass (horizontal axes). We notice that the priority UOP scenario may reach a sensitivity that would directly detect dark matter in the Solar System for masses below $\sim 10^{-23}$\,eV. While these low masses are already in tension with other astrophysical observations (see, \eg, Refs.~\cite{Bar:2018acw,Ferreira:2020fam}), this is a very interesting result since, as compared to other bounds, it would be based on \emph{direct} sensitivity to the ULDM field. 

Regarding direct coupling, from Eq.~\eqref{eq:direct} and the bounds on the gravitational effect from Fig.~\ref{fig:DM}, it is easy to understand the constraining power on $\Lambda_2$ for different masses. 
In the upper panel of Fig.~\ref{fig:DM} we show the equivalent $h_c$ predicted for two values of $\Lambda_2$: one corresponding to the typical minimum value UOP could constrain (around $\Lambda_2=10^{-18}\,\rm GeV^{-1}$) and another one that improves on the best bounds from Cassini ($\Lambda_2=10^{-16}\,\rm GeV^{-1}$). We see clearly that in both cases, even the baseline UOP will produce the most stringent results for ULDM in a large range of masses. To make this more explicit, in Fig.~\ref{fig:DM}, lower panel, we show the constraints that can be achieved on $\Lambda_2$ as a function of the DM mass for different configurations
and compare them to \textit{Cassini} and PTA~\citep{2017PhRvL.118z1102B} data (see also Ref.~\citep{Blas:2019hxz} and the recent Refs.~\citep{Kus:2024vpa,Smarra:2024kvv} for related bounds). It is clear that the UOP will enormously extend the searches for these models, with the potential to generate the first direct detection of dark matter.

Finally, we also show in Fig.~\ref{fig:DM} the equivalent $h_c$ from the effect from Eq.~\eqref{eq:shift} representing axion-like particles. Recall that this effect also arises from an oscillating part of the ULDM with frequency $f_{\rm DM}$. We have used a value of the coupling $g_{a \gamma\gamma}$ characteristic of current bounds at the relevant masses, see Ref.~\cite{AxionLimits}. From this plot, we conclude that the priority or optimistic cases of the UOP mission would also have a unique legacy in the search for axion-like particles.

\section{Beyond the Baseline: Reducing the Allan deviation}
\label{sec:noise}
\subsection{Noise sources}
Predicting realistic improvements in the Allan deviation requires considering recent advancements in a wide range of technologies. Here we base some qualitative considerations on the thorough discussion in Ref.~\cite{2006armstrong}. {All quoted noise magnitudes are in terms of the Allan deviation, and are therefore dimensionless.} 

\textit{Cassini}-era noise was dominated by 3 major components; the antenna mechanical noise, the plasma interference noise, and the tropospheric scintillation noise. The first is simply a mechanical limitation, which can be addressed by implementing a complementary smaller and stiffer antenna in combination with the main dish, as well as via three-point antenna calibration, which carries the potential to reduce $\sigma_{\rm A}$ of Ka-band tracking missions by one order of magnitude.\footnote{Private conversation
with Sami W. Asmar, NASA JPL.}
The noise of plasma scintillation, the variation of the refractive index of the interplanetary medium, is localised around $3\times10^{-3}$ Hz and then steadily drops down for lower frequencies~\citep{plasma}. It is astrophysical in origin, and therefore cannot be easily addressed with technological improvements. One possibility is to prioritise tracking measurements at optimal Sun-Earth-spacecraft configuration, as well as by upgrading the Doppler link to higher frequencies, as discussed in Section \ref{subsec:optical}.

Tropospheric noise is thought to dominate the noise PSD at frequencies below $10^{-4}$ Hz. Since the \textit{Cassini} launch, corrective measures based on water-vapor-radiometers have been shown to be able to reduce the Allan deviation up to a factor of 10, down to $1.5 \times 10^{-15} - 3 \times 10^{-15}$, {allowing for the GW constraints placed by the \textit{Cassini} 40 day observation run~\cite{2006armstrong}}. Furthermore, drastic improvements in tropospheric noise would be achievable by tracking the spacecraft with high-altitude facilities, either ground-based or with balloons. Finally, an alternative opportunity to reduce tropospheric noise would be using multiple measurement points via a radio telescope array~\citep{tropo}.

Glitches and unmodelled accelerations, \eg, from solar winds or imprecisions in the planetary ephemeris, may also leave residual imprints on the Doppler time series. However, these outliers will not influence the coherent stacking of the GW signal in the experiment. A more quantitative treatment of their influence can be addressed by a more detailed simulation of the mission's trajectory, once the satellite specifications are known. Connected to this, one must also consider the noise from asteroids in the Solar system. A simple estimate using Brownian-motion argument~\cite{Loeb:2024ekw} suggests a noise level in the solar system barycenter of order $10^{-15}$ in the 10~nHz range due to asteroids smaller than 80~km. Furthermore, the results in Ref.~\cite{Fedderke:2020yfy} indicate a noise level lower than $10^{-18}$ for the potential generated by asteroids on a satellite at $\sim$AU from the Sun (with typical frequencies around and below $\mu$Hz), which would be irrelevant to the scientific outcome we outlined. However, a more detailed analysis confirming this expectation and considering a realistic mission trajectory and possible mitigation measures using updated asteroid ephemerides (see Ref.~\cite{NANOGrav:2020tig} for the current strategy followed by NANOGrav) seems opportune. {We reserve a more thorough analysis of different noise sources in the low frequency part of the noise PSD for future work.}

\subsection{Breakthrough: Optical links?}
\label{subsec:optical}

A true breakthrough in improving the Allan deviation to $\lesssim10^{-15}$ could be achieved by upgrading the Doppler link to optical frequencies. Here we briefly review theoretical limits and recent experiments. There are three distinct techniques in optical ranging.
\begin{itemize}
  \item The passive-reflector method has reflecting cubes
    on the target, which reflect light back to the source. This has
    been very successful for lunar ranging~\citep{2019JGeod..93.2195M}
    but (as we will see below) is not feasible at interplanetary
    distances.
  \item The synchronous transponder method has the target receive a
    signal and then send another signal back with a fixed delay. This
    has been recently demonstrated with the Hayabusa2 mission to
    $\sim15$ lunar distances~\citep{2023AdSpR..71.4196N}, with ground
    observations even during the day.
  \item An asynchronous transponder ranging both sides sends signals at
    pre-designated intervals, and the ranging is computed indirectly~\citep{2002JGeo...34..551D}.  The Deep Space Optical
    Communications (DSOC) system on the Psyche mission~\citep{2021SPIE11721E..0SP,54760_2021} is an example.
\end{itemize}

We can get an idea of what is and is not feasible using
order-of-magnitude arguments. Similar arguments appear in
\cite{2023Photo..10...98D}. Consider a signal at wavelength $\lambda$
from a transmitting telescope of area $A_T$ to a receiving telescope
of area $A_R$ at distance $R$. A diffraction-limited beam will have a
solid angle of $\sim\lambda^2/A_T$. Of the photons transmitted from
$A_T$, the fraction received at $A_R$ will be $\sim A_R/R^2$ divided
by the solid angle. Thus a fraction
\begin{equation} \label{eq:phfrac}
  \frac{A_T A_R}{\lambda^2 R^2}
\end{equation}
will be received. For a transmitting power $P$, the rate of photons
received is then
\begin{equation}
  \frac{P}{hc\lambda} \frac{A_T A_R}{R^2}.
\end{equation}
The maximum data rate for a given power and area is thus
$\propto\lambda^{-1}$.

For some orders of magnitude, let us put $\lambda=\SI{1e-6}{\metre}$,
$A_T=A_R=\SI{1}{\metre^2}$ and $R=\SI{3e12}{\metre}$. This makes
the photon fraction $\sim10^{-13}$ and the photon rate
$\sim\SI{5e4}{\watt^{-1}}$.

With a passive reflector the photon fraction (\ref{eq:phfrac}) gets squared through the round trip. At lunar
distances ($R\simeq\SI{4e8}{\metre}$) this is tolerable, but at
interplanetary distances, it would become hopeless. Hence an active
system is needed.

Ref.~\cite{2020RSPTA.37890488S} discusses optical telecom prospects for
ice-giant missions briefly, noting the positives and negatives, which also follow from the above arguments.
\begin{itemize}
\item[$+$] The data rates can be orders of magnitude higher.
 Ref.~\cite{2014SPIE.8971E..0WS} demonstrated uplink and downlink rates of
  19 and 38~Mbps respectively from lunar distances. The Psyche mission to the asteroid belt was designed for rates of 250~Mbps from deep space~\citep{2019SPIE10978E..09B}, and a demonstration of the data rate (showing a cat pursuing a laser pointer) was widely reported in the media and social media. 
\item[$-$] The narrowness of the optical beam requires accurate
  pointing, including the effect of light travel time.
\item[$-$] To be weather-proof, the system requires redundant ground
  stations (or balloon or orbiting stations).
\end{itemize}

Laboratory experiments of asynchronous ranging
\citep{2010SPIE.7587E..0AB} give precisions of better than 1~mm, which
amounts to $3\times10^{-16}$ of the Uranus distance. Reference
\cite{2019JGeod..93.2405D} discuss the error budget for interplanetary
ranging, and conclude that there is potential for (sub-)mm range
accuracy. That is, accuracies even better than the optimistic scenario considered earlier appear possible with optical ranging.  Additional considerations are discussed Ref.~\cite{2024mcquinn}.


\section{Conclusion: Mission Outlook}
\label{sec:conclusion}

We have demonstrated that the prospective flagship mission to Uranus (UOP) has rich scientific opportunities that go beyond its nominal planetary science goals. Doppler tracking data may be accumulated over the entire interplanetary cruise phase, and subsequently stacked according to the methodology showcased in Section \ref{subsec:stack}. This results in an improved capacity to detect imprints from deterministic GWs from individual SMBH binaries (Section \ref{S:smbh}), astrophysical 
 as well as early universe stochastic GW backgrounds (Sections \ref{S:smbh} and \ref{sec:earlyuniverse}), and finally gravitational and non-gravitational couplings between the spacecraft and dark matter (Section \ref{sec:uldm}). Such varied and timely prospects would greatly expand the scientific yield of the mission, broadening its relevancy to communities beyond the planetary sciences. So the question beckons: what is required to actually achieve what we have shown to be possible in this paper? We note that our analysis of the mission's capacity to distinguish various signals should be repeated with more sophisticated noise models, including, \eg, non-Gaussianities and the time dependence of different noise contributions along the cruise phase. {This aspect will be more significant than for the \textit{Cassini} 40 day experiment, due to the $\sim$100 times longer duration and the large span of the trajectory from $\sim 5$ AU to $\sim$20 AU.} A more thorough treatment of the mission's trajectory is also crucial, in particular considering the influence of unmodelled accelerations due to uncertainties in the masses of planets, asteroids, and solar radiation pressure. The latter aspect has been estimated in, \eg, Ref.~\cite{2022fedderke} for an application of ranging with asteroids, and would potentially result in noise strains of order few $10^{-14}$ at $10^{-7}$ Hz for a rough estimate of the UOP satellite mass and cross section (private communication with the authors). Additionally, a more sophisticated data analysis pipeline should take into account the possibility of multiple signals to be present at the same time, including the possibility of stochastic backgrounds overshadowing individual signals. Nevertheless, taking these limitations at face value, we have demonstrated how two crucial factors determine the magnitude of the achievable scientific goals. They are:
\begin{itemize}
    \item \textit{The total amount of tracking data} collected during the cruise phase, which could range from a few data points to near to constant tracking.
    \item \textit{The technological feasibility of reducing the Allan deviation} with respect to its baseline value of $\sim 10^{-14}$ at $10\,$s, as observed for the \textit{Cassini} mission data more than three decades ago, in 1992.
\end{itemize}
We maintain that these two points should become part of the mission requirements, in light of the numerous scientific opportunities demonstrated in this work. With them, the UOP mission has the potential for groundbreaking discoveries about black holes living in galactic centers, the study of particle physics beyond the standard model and the reach of current particle accelerators, and the nature of dark matter.

\begin{acknowledgments}
%
This research was supported by the International Space Science Institute (ISSI) in Bern, through ISSI International Team project 551 ({\em Future Missions to Uranus and Neptune: Prospects for Non-Planetary Science}).
L.Z. and D.O. acknowledge support from ERC Starting Grant No. 121817–BlackHoleMergs.
D.J.D. and D.O. received funding from the Danish Independent Research Fund through Sapere Aude Starting Grant No. 121587. A.D. acknowledges support from NSF grant AST-2319441.
We thank Mark Hofstadter, Joseph Lazio, Daniele Durante, and Frenceso Iacovelli for useful discussions. D.S. thanks Timothée Schaeffer for continuing moral support and inspiration. L.Z. also acknowledges G.M.'s curiosity. 
The research leading to these results has received funding from the Spanish Ministry of Science and Innovation (PID2020-115845GB-I00/AEI/10.13039/501100011033).
IFAE is partially funded by the CERCA program of the Generalitat de Catalunya. D.B. acknowledges the support from the Departament de Recerca i Universitats de la Generalitat de Catalunya al Grup de Recerca i Universitats from Generalitat de Catalunya to the Grup de Recerca 00649 (Codi: 2021 SGR 00649).
A.C.J. is supported by the Science and Technology Facilities Council (STFC) through the UKRI Quantum Technologies for Fundamental Physics Programme [grant number ST/T005904/1].
This work was partly enabled by the UCL Cosmoparticle Initiative.
 
\end{acknowledgments}

\appendix
\section{White noise Bayesian sensitivity curve}
\begin{figure*}
    \centering
    \includegraphics[width = 0.32\textwidth]{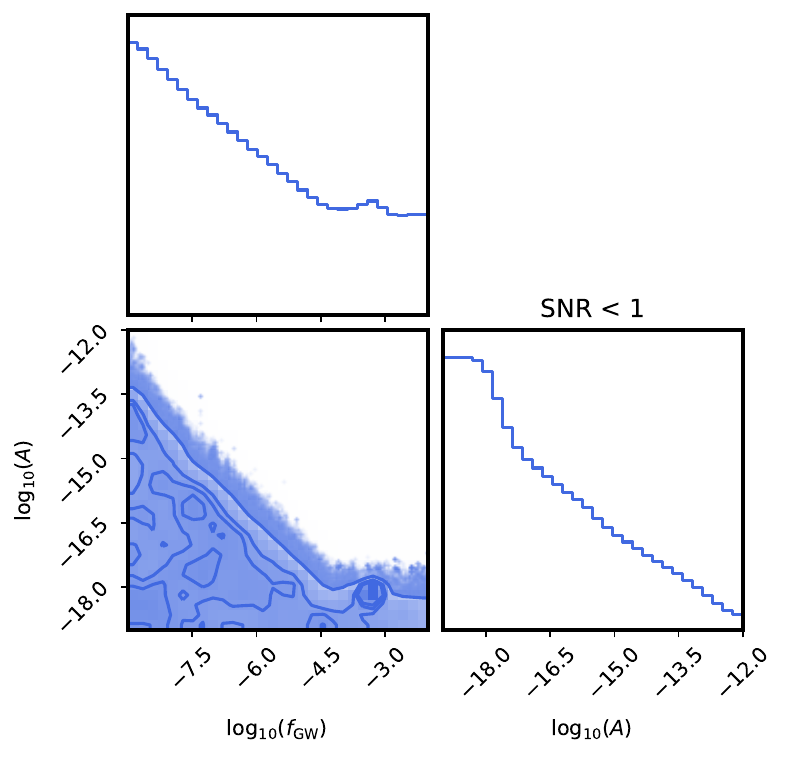}
    \includegraphics[width = 0.32\textwidth]{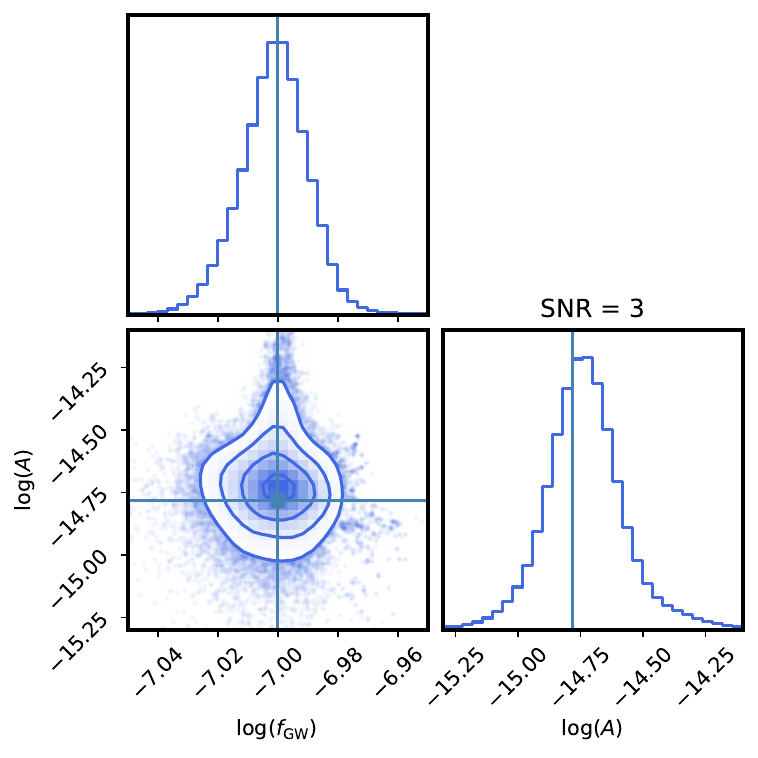}
    \includegraphics[width = 0.32\textwidth]{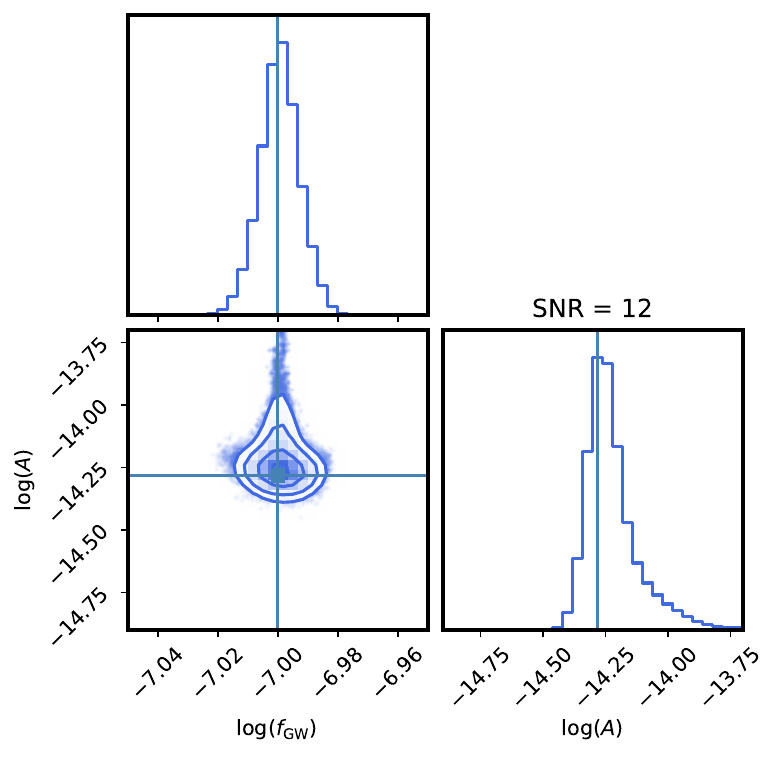}
    \caption{We show the two-dimensional marginalised posterior distributions in amplitude and frequency for a set of three different injected GW amplitudes, corresponding to an SNR of 0, 3, and 12 from left to right. Here the SNR is calculated according to Eq.~\eqref{eq:SNRgw} with respect to the threshold amplitude shown in Fig.~\ref{fig:1sigma}. As we can see, increasing SNR produces better parameter estimation in both frequency and amplitude of the GW, where sources with SNR=3 can already be constrained. The upper bound on the GW amplitude suffers from a degeneracy with the source's position in the sky (not shown) and therefore stops improving significantly after an SNR$\sim 8$ for this particular mission trajectory.}
    \label{fig:mcmc1}
\end{figure*}
In this Appendix, we explore an additional way to define the mission's sensitivity curve that relies on Bayesian techniques, such that it may better represent the actual parameter estimation procedure. The methodology here developed is limited to white Gaussian noise, and thus only serves as an indicative evidence that the data analysis problem of Doppler ranging is solvable.

A possible way to define a sensitivity curve is to sample the likelihood of a tracking time series that does not contain any GW signal. We are looking to evaluate the posterior probability of a GW with parameters $ \left[\hat{\boldsymbol{k}}, f_{\scriptscriptstyle\rm GW}, A,\phi_0 \right]$ to be present in a given realisation of $y_2^{\rm{tot}}$. Or in other words, we wish to evaluate a likelihood function $\mathcal{L}$ of the form:
\begin{align}
    \log\left(\mathcal{L} \right) \propto - \sum\left(y_2^{\rm{tot}} - y_2^{\rm{trial}} \right)^2,
\end{align}
where $y_2^{\rm{trial}}$ is a trial GW response. To this end, we employ a Markov Chain-Monte Carlo (MCMC) framework, where we use the Monte Carlo sampling package \texttt{emcee}~\citep{emcee} running 32 parallel walkers in order to recover the posteriors of the GW parameters $\left[\hat{\boldsymbol{k}}, f_{\scriptscriptstyle\rm GW}, A,\phi_0 \right]$. 

The MCMC walkers will be able to freely sample regions in which the trial GW amplitude $A^{\rm{trial}}$ is well below the noise. However, they will be strongly discouraged to sample large values of $A^{\rm{trial}}$, since they are inconsistent with the absence of a signal. Thus, the posterior probability for the GW amplitude $A$ will sharply decrease above a certain threshold $A_s$:
\begin{align}
    \mathcal{P}(A) \sim \begin{cases} {\text{const.}} & {\text{for }} A < A_s\left( \hat{\boldsymbol{k}}, f_{\scriptscriptstyle\rm GW},\phi_0\right),\\
     0 & {\text{for }} A > A_s\left(\hat{\boldsymbol{k}}, f_{\scriptscriptstyle\rm GW},\phi_0 \right),
    \end{cases}
\end{align}
After marginalising over the initial phase $\phi_0$ and wave vector $\hat{\boldsymbol{k}}$, we can specify the threshold to the desired confidence level and obtain a curve $A_s(f_{\scriptscriptstyle\rm{GW}})$, purely as a function of GW frequency. An example of this procedure is visualised in the left panel of Fig.~\ref{fig:mcmc1}, where one can clearly see a transition between allowed and disallowed regions of parameter space in the 2-dimensional posteriors of the GW frequency and amplitude.

\begin{figure}
    \centering
    \includegraphics[width=\columnwidth]{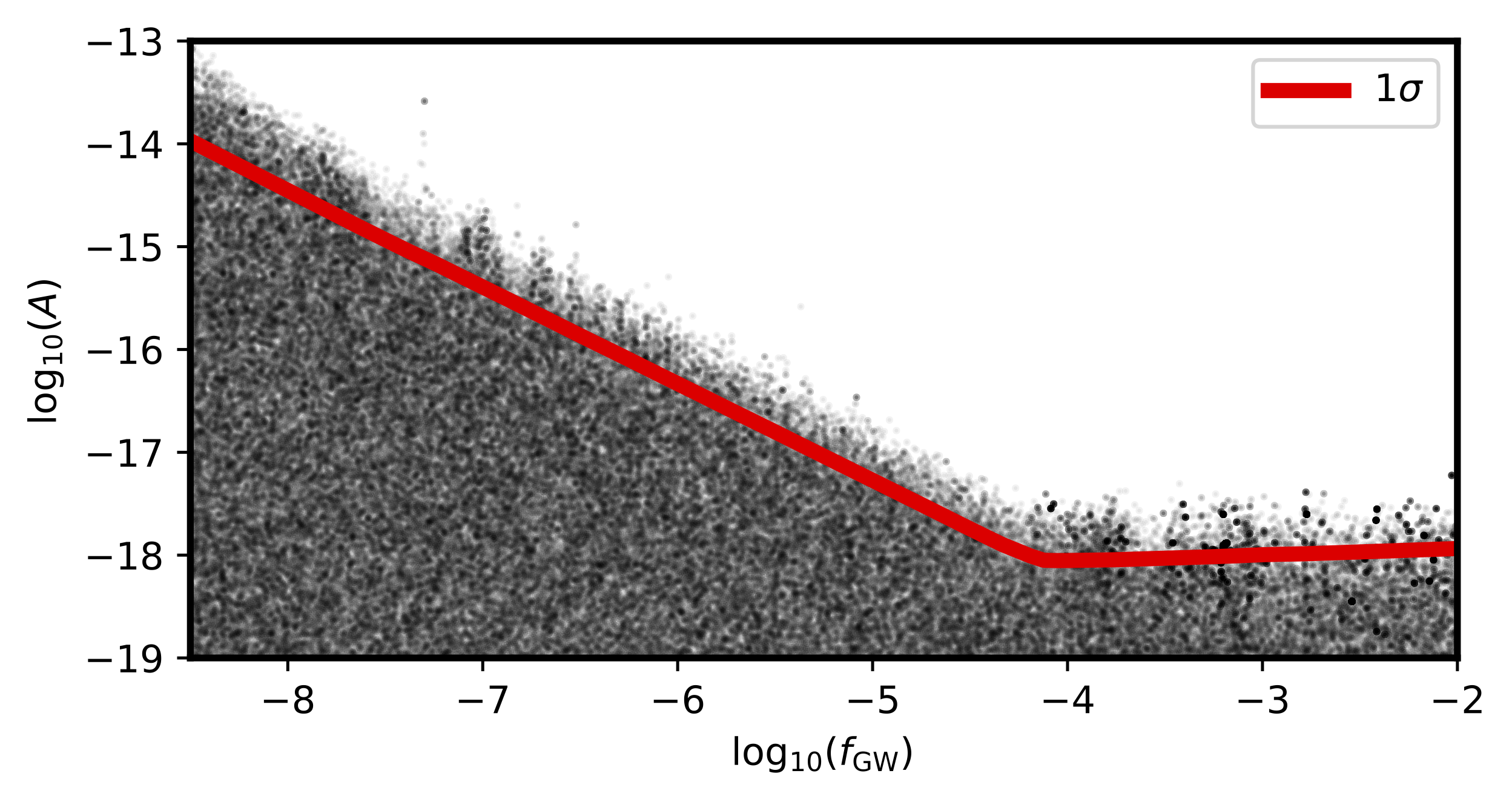}
    \includegraphics[width=\columnwidth]{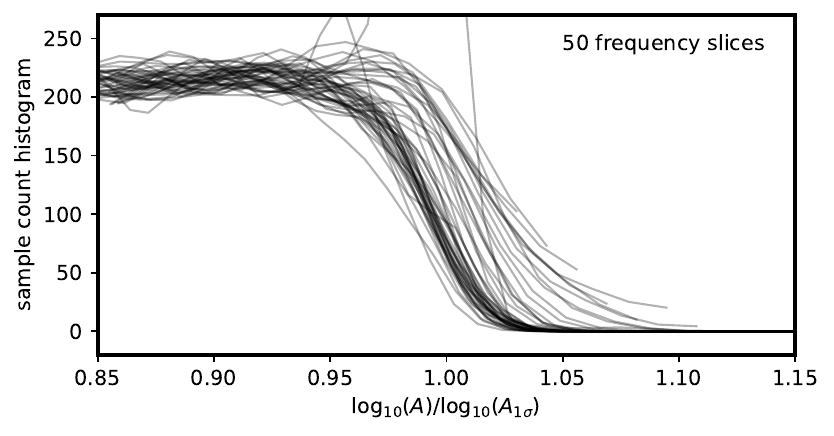}
    \caption{In the top panel we show the two-dimensional marginalised posterior samples in amplitude and frequency for a run with no injected signal. We highlight the threshold at which the absence of a signal is excluded at $1\sigma$ confidence, which is well fit by a broken power law. In the lower panel, we showcase the sample count in 50 frequency bins, as a function of amplitude. We normalise the latter with the $1\sigma$ threshold.}
\label{fig:1sigma}
\end{figure}

As seen in the bottom panel of Fig.~\ref{fig:1sigma}, the boundary between the allowed and disallowed regions is closely approximated by a half-Gaussian. This allows us to easily find the $1\sigma$ confidence threshold, above which signals are distinguishable from the null hypothesis. Note that this does not necessarily imply that the individual signal parameters are well constrained, only that the overall signal is distinguishable from its absence. This corresponds to the analogous definition of a sensitivity curve for other GW detectors (see, \eg, Ref.~\cite{2019robson}). For the purposes of defining a simple sensitivity curve, we fit the one sigma boundary with a broken power law, shown in the top panel of Fig.~\ref{fig:1sigma}.

The value of the $1\sigma$ threshold $A_s$ depends on the noise PSD of the tracking runs, the specified mission trajectory, as well as the total number of tracking runs $N_{\rm t}$. By performing several numerical MCMC tests, we find, as one would expect, that its overall normalisation scales linearly with the Allan deviation and scales as the square root of the total number of samples:
\begin{align}
    A_s \sim \sigma_{\rm A}(\tau) \times \frac{\sqrt{\tau}}{ \sqrt{N_{\rm t}}}.
\end{align}
The effect of the spacecraft's trajectory is also folded into the specific realisation of $A_{s}$, as it modulates the GW response function $y_2^{\scriptscriptstyle\rm{GW}}$ over the 10-year observation window. In particular the relative orientation of the wave vector $\hat{\boldsymbol{k}}$ and the link vector $\hat{\boldsymbol{n}}$ can change significantly for elliptical trajectories. The presence of curvature in the trajectory is crucial, as it allows one to break the sky localisation degeneracy that would be present for a single, straight detector arm.

To reiterate, the recovered $1\sigma$ threshold $A_s$, which corresponds to the definition of a sensitivity curve, results from distributing $N_{\rm t}$ tracking runs over the entire mission duration of $T = 10$ years. For a monochromatic signal, this results in tracking a total of $n\sim (f_{\scriptscriptstyle\rm{GW}}\times N_{\rm t} \times 8\,\rm hour)$ cycles. This is reminiscent of matched filtering techniques, which enable one to compare the sensitivity curves of GW observatories to the characteristic strain of a signal, rather than its instantaneous amplitude~\citep[see, \eg,][]{2019robson}. Thus, we adjust the $1\sigma$ detection threshold by heuristically factoring out the number of detected cycles:
\begin{align}
    h_{\rm{sens}}(f_{\scriptscriptstyle\rm{GW}}) = A_s(f_{\scriptscriptstyle\rm{GW}}) \times\left( f_{\scriptscriptstyle\rm{GW}}\times N_{\rm t} \times 8 \, {\rm{hr}} \right)^{1/2},
\end{align}
where we define $h_{\rm{sens}}(f_{\scriptscriptstyle\rm{GW}})$ the GW sensitivity curve of the mission, given a trajectory, an Allan deviation and a set of tracking day timestamps $t_n$:
\begin{align}
    h_{\rm{sens}}(f_{\scriptscriptstyle\rm{GW}}) = h_{\rm{sens}}(f_{\scriptscriptstyle\rm{GW}}; \sigma_{\rm A} (10\, {\rm{s})}, t_n).
\end{align}
Then, we can also define the SNR of a GW detected with Doppler tracking in complete analogy with other gravitational wave detectors:
\begin{align}
    {\rm{SNR}} = \sqrt{4 \times \int \frac{h_{\rm c}(f^\prime)^2}{h_{\rm{sens}}(f^\prime)^2 f^\prime} {\rm d} f^\prime},
\end{align}
where $h_{\rm c}$ is the GW's characteristic strain, and the factor 4 comes from the normalisation of the one-sided-power spectrum. Here, an SNR of $1$ corresponds to a barely detectable GW that can be distinguished from the absence of signal with $1\sigma$ confidence. As shown in Fig.~\ref{fig:mcmc1}, higher SNRs result in both better parameter estimation and in reduced false alarm rates, also in complete analogy with GW detectors. We find that the frequency and amplitude of sources with an SNR$\sim3$ can be recovered within $\sim 0.02$ and $\sim 0.5$ dex, respectively. The bounds improve by a factor $\sim 3$ for SNR $=12$ sources, though a residual degeneracy between the source's amplitude and optimal localisation in the sky becomes manifest, inducing a skew in the recovered posterior distribution.
\section{Sky dependence}
\label{app2}
\begin{figure}
    \centering
\includegraphics[width=0.99\linewidth]{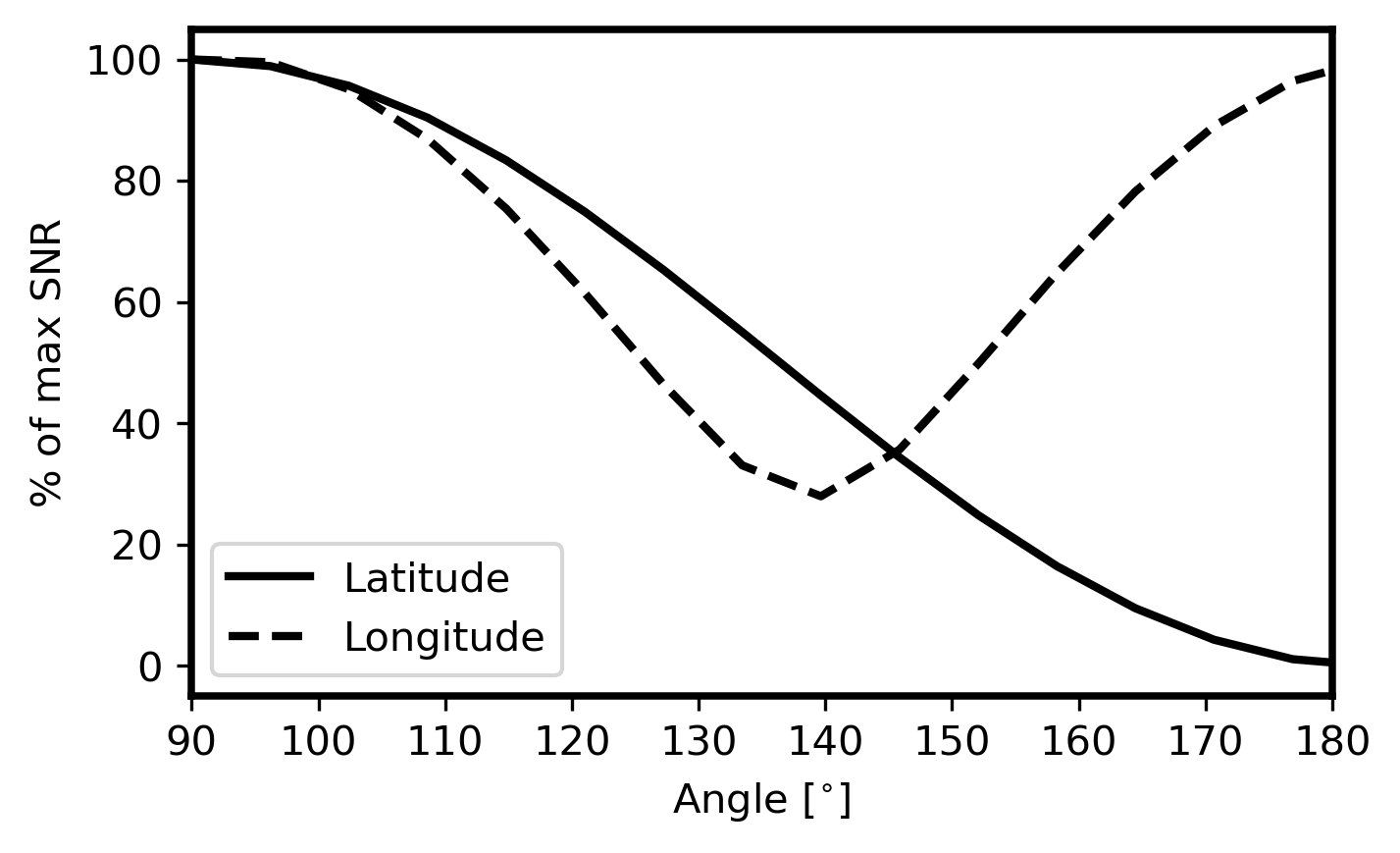}
    \caption{Sky dependence of the SNR for GW sources located at a given latitude and longitude with respect to the orbital plane and initial vector between the spacecraft and the Earth. Note the lack of a "blind-spot" in the longitude, which is due to the curvature in the orbit.}
    \label{fig:placeholder}
\end{figure}

The typical sky-averaging of the GW strain is modified by the variable detector geometry of the UOP. We include a plot for visualisation purpuses in Fig. \ref{fig:placeholder}.

\bibliography{issi}

\bibliographystyle{apj}

\end{document}